\documentclass[useAMS,usenatbib]{mn2e}
\def\aap{AA}

\def\mnras{MNRAS}
\def\apj{ApJ}

\def\nat{Nat}

\def\prd{PhysRevD}
\def\jcap{JCAP}
\def\physrep{Phys. Rept.}
\def\araa{Annual Review of Astronomy and Astrophysics}

\def\dgemail{gilmanda@ucla.edu}

\usepackage[titletoc]{appendix}
\usepackage{graphicx}
\usepackage{float}
\usepackage{amssymb}
\usepackage{amsfonts}
\usepackage{amsmath} 
\usepackage{color}
\usepackage{wrapfig}
\usepackage{hyperref}
\usepackage{algorithm}
\usepackage{algpseudocode}

\usepackage{natbib}

\def\msub{{\bf{m}_{\rm{sub}}}}
\def\qsub{{\bf{q}_{\rm{sub}}}}
\def\qmac{{\bf{q}_{\rm{mac}}}}
\def\qmacn{{\bf{q}_{\rm{mac(n)}}}}
\def\data{{\bf{d}_{\rm{n}}}}
\def\Data{{\bf{D}}} 

\def\fobs{{\bf{f_n}}}
\def\fmod{{\bf{f}_{n}^{\ \prime}}}
\def\dlinmod{{\bf{d}_{tx(n)}^{\ \prime}}}
\def\dlin{{\bf{d}_{tx(n)}}}

\def\qmacs{{\bf{q}_{\rm{mac(n)}}^{\ \star}}}
\def\fmods{{\bf{f_n}^{\star}}}

\def\msun{{\rm{M}_{\odot}}}
\def\fobsi{{f_{n\left(i\right)}}}
\def\fmodsi{{f_{n\left(i\right)}^{\ \star}}}

\title{Probing the nature of dark matter by forward modeling flux ratios in strong gravitational lenses} 
\author[Gilman et al.]{\parbox{\textwidth}{
		Daniel Gilman$^{1}$\thanks{\dgemail}, Simon Birrer$^{1}$, Tommaso Treu$^{1}$, Charles R. Keeton$^{2}$, Anna Nierenberg $^{3}$}
	\\
	\\
	\parbox{\textwidth}{
		$^{1}$Department of Physics and Astronomy, University of California,
		Los Angeles, CA 90095, USA\\
		$^{2}$ Department of Physics and Astronomy, Rutgers, the State University of New Jersey, 136 Frelinghuysen Road, Piscataway, NJ 08854\\
		$^{3}$ Department of Physics and Astronomy, University of California Irvine, Irvine CA, 92697-4575}
}

\begin{document}
	
	\voffset-.6in
	
	\date{Accepted . Received }
	
	\pagerange{\pageref{firstpage}--\pageref{lastpage}} 
	
	\maketitle
	
	\label{firstpage}
	
	\begin{abstract}
		The free streaming length of dark matter particles determines the abundance of structure on sub-galactic scales. We present a statistical technique, amendable to any parameterization of subhalo density profile and mass function, to probe dark matter on these scales with quadrupole image lenses. We consider a warm dark matter particle with a mass function characterized by a normalization and free streaming scale $m_{\rm{hm}}$. We forecast bounds on dark matter warmth for 120-180 lenses, attainable with future surveys, at typical lens (source) redshifts of 0.5 (1.5) in early-type galaxies with velocity dispersions of 220-270 km/sec. We demonstrate that limits on $m_{\rm{hm}}$ deteriorate rapidly with increasing uncertainty in image fluxes, underscoring the importance of precise measurements and accurate lens models. For our forecasts, we assume the deflectors in the lens sample do not exhibit complex morphologies, so we neglect systematic errors in their modeling. Omitting the additional signal from line of sight halos, our constraints underestimate the true power of the method. Assuming cold dark matter, for a low normalization, corresponding the destruction of all subhalos within the host scale radius, we forecast $2\sigma$ bounds on $m_{\rm{hm}}$ (thermal relic mass) of $10^{7.5} \ (5.0)$, $10^{8} \ (3.6)$, and $10^{8.5} \ (2.7) \ M_{\odot} \left(\rm{keV}\right)$ for flux errors of $2\%$, $4\%$, and $8\%$. With a higher normalization, these constraints improve to $10^{7.2} \ (6.6)$, $10^{7.5} \ (5.3) $, and $10^{7.8} \ (4.3) \ M_{\odot} \left(\rm{keV}\right)$ with 120 systems. We are also able to measure the normalization of the mass function, which has implications for baryonic feedback models and tidal stripping.
	\end{abstract}
	
	\section{Introduction}
	Dark matter models make testable predictions regarding the abundance and mass profiles of substructure in galactic dark matter halos. In the standard $\Lambda$CDM picture, structure grows bottom-up at practically all length scales \citep{Schneider++13}, resulting in a scale-free mass function for subhalos \citep{Spr++08,Gao++12,Fiacconi++16}, and density profiles fit by the Navarro Frenk and White (NFW) profile \citep{NFW97}. In contrast, in warm dark matter (WDM) models, free streaming washes out small density perturbations, resulting in a paucity of structure below a certain scale, which depends on the free streaming scale of the dark matter particle(s) \citep{Schneider++12,Pullen++14,Viel13,Lovell++16,BoseHellwing++16,Menci++16}. Thermal relics and sterile neutrinos represent two WDM candidates \citep{Kusenko09,Abazajian17}, and under certain assumptions produce similar mass functions. Self-interacting dark matter alters the density profiles of individual subhalos, producing cores rather than cusps in the center of halos \citep{Schneider++16,Vogelsberger++16,Kamada++17}.
	
	These alternative models to $\Lambda$CDM have gained traction, motivated by apparent failures of the $\Lambda$CDM picture on small scales \citep[see][]{BullockBK17}. For example, by invoking WDM one explains the non-detection of low-mass satellites in the Milky Way, the ``missing satellites problem'' \citep{Kly++99}, by reducing the expected number of subhalos. In another example, rotation curves of satellite galaxies imply shallower-than-isothermal inner density profiles \citep{deBlok10}, characteristic of the mass profiles associated with self interacting dark matter. 
	
	These apparent anomalies are based on observations of luminous structure and rely on assumptions about the connection between baryons and dark matter halos. Under different combinations of models for star formation and kinematic data in satellites, some small-scale challenges to the $\Lambda$CDM picture can be resolved \citep[e.g.][]{KIM++2017}. Unfortunately, it is difficult to measure kinematics of low mass galaxies owing to the small number of stars that can be used for this purpose. This leads to large uncertainties in the inferred halo masses and densities. Different models for tidal disruption and baryonic feedback \citep{DespVeg16,GK++17}, differing luminosity functions for dark subhalos \citep{Nie++16}, and large scatter in stellar-mass-halo-mass relations below $10^8 \msun$ \citep{Munshi++17} complicate constraints on the nature of dark matter. An independent and direct probe, which does not rely on assumptions regarding the physics of star formation in low mass galaxies, is needed to disentangle the physics of baryons and the nature of dark matter. 
	
	Gravitational lensing offers a direct probe of dark substructure below halo masses of $10^8 \msun$. Lensing relates a set of three observables - time delays, positions, and magnifications - to the gravitational potential along a path traversed by light emitted by a background source. As the observables depend only on the gravitational potential of the deflector and the potential along the line of sight, lensing offers a tool to study dark matter substructure directly, without relying on baryons as tracers.
	
	Various techniques employ strong lensing as a probe of dark matter structure. When the light from a spatially extended background source is warped by a foreground deflector into a highly magnified arc, substructure can distort the arc. By simultaneously modeling the mass distribution in the lens plane and reconstructing the luminosity distribution of the background source, one can infer the mass of a perturber and constrain the subhalo mass function \citep{Koo05,Vegetti:2010p30241,Veg++12,Veg++14,Li++16,Hezaveh++16c,Birrer++17,Vegetti++18}.
	
	Flux ratios in lensed quasars offer an alternative probe of dark substructure \citep{MaoSchneider98,MetcalfMadau01,D+K02,Chiba02}. The sensitivity of this observable derives from the compact size of a quasar and the fact that lensing magnifications can be perturbed by a subhalo whose deflection angle is comparable to or larger than the source size \citep{D+K06b}. From a theoretical standpoint, several forecasting studies \citep[e.g.][]{Xu++09,Xu++12,Graus++17} use N-body simulations to anticipate the lensing signal from CDM substructures. From an observational and lens modeling point of view, given observed flux ratios, one can add a subhalo to the lens model and vary its properties to infer the mass \citep[e.g.,][]{FadelyKeeton12,MacLeod++13,Nie++14} or rule out the presence of substructure near the images \citep[e.g.,][]{Nierenberg++17}. This lens modeling technique, and the direct detection of subhalos via gravitational imaging, comprise a class of observations that yield constraints on individual substructures.
	
	Recently, authors have called attention to a potential bias present in flux ratios, wherein the morphology of deflector in the lens plane produces flux ratio anomalies reminiscent of those induced by substructure. \citet{Gilman++17} showed that realistic deflectors with a luminous mass component drawn from HST images of nearby galaxies occasionally produce flux ratio anomalies with respect to a simplified smooth lens model. \citet{Hsueh2++17} performed a similar analysis with galaxies produced in the Illustris simulation \citep{VogelsbergerIllustris14}, and reached similar conclusions. In two observed lenses, \citet{Hsueh++16,Hsueh++17} argued that a disk component in the deflector can explain observed flux ratio anomalies. These effects may contribute to frequency of observed flux ratio anomalies in strong lenses, which occur more frequently than one would expect in a CDM scenario, as pointed out by \citet{Xuetal2015}. Among non-dark matter sources of flux ratio anomaly, microlensing can also induce drastic fluctuations in image magnifications, although this effect can be mitigated by using flux ratios measured from the more spatially extended narrow-line region of a background quasar \citep{MoustakasMetcalf03,Nie++14,Nie++16}.
	
	In the context of dark matter, it is important to note that models predict large ensembles of dark subhalos, which could act together to affect a lensed image. In contrast to single subhalo models, other methods attempt to probe the collective impact of numerous perturbers whose individual effects are not statistically significant, but which together produce a measurable signal \citep{D+K02,FadelyKeeton12,Hezaveh++16b,CyrRacine++16,Daylan++17,C+K17,Birrer++17}. Since these methods do not require high significance detections of individual perturbers, they extract information from a larger area around Einstein rings. 
	
	In an example, \citet{Birrer++17} quantify substructure in the lens RX J1131$-$1231 by modeling surface brightness anomalies detected in HST imaging data. Through a forward modeling approach that relies on generating an extensive suite of realistic simulations, they are able to constrain models of dark matter statistically. They could rule out WDM mass functions with thermal relics below the mass of 2 keV at $2\sigma$.
	
	In this work, we present a statistical method that utilizes the flux ratios from an ensemble of multiply imaged quasars to distinguish between dark matter models. Our technique takes as input data from a sample of strong lens systems and a prescription for rendering substructure realizations for a dark matter theory, and returns posterior probability distributions for the parameters describing the substructure population. We use the technique of Approximate Bayesian Computing \citep[ABC;][]{Rubin1984ABC}, also applied in \citet{Birrer++17}, which enables an application of Bayesian statistics to the problem of substructure lensing. In our framework, we are able to efficiently explore the parameter space spanned by dark matter models with different predictions regarding the nature of substructure without explicitly computing a likelihood function, which in substructure lensing is a computationally and analytically daunting task. ABC permits one to circumvent calculation of formidable likelihoods through the use of summary statistics, which quantify agreement between an observation and data computed in a forward model. Our method also naturally accommodates joint inference from multiple strong lens systems. The method can be applied to any parameterization of dark substructure, provided one specifies the mass function, spatial distribution, and density profile of individual subhalos. Since the method relies on flux ratio statistics rendered in a forward model, accurate lens models and control over systematic errors in flux ratios are crucial for attaining robust constraints. Finally, by omitting line of sight substructure that is expected to contribute significantly to flux ratio anomalies \citep{Chen++03,Inoue++16,Despali++17}, we do not capture the full information content of each lens, so our results can be interpreted as conservative theoretical bounds.
	
	We present the formalism of our method and illustrate its general capabilities via a case study in which we distinguish between two simplified dark matter models. We consider a subhalo mass function with variable normalization and damping below a free-streaming scale, and provide forecasts for the constraints afforded under different flux precisions with up to 180 systems, which is a sample size attainable with future surveys such as Euclid, LSST and WFIRST \citep{O+M10}. With our forward modeling framework, we forecast the possible constraints on a WDM subhalo mass function using flux ratios, and quantify how these constraints scale with the number of lenses, the uncertainty in fluxes, and the overall normalization of the mass function.
	
	This paper is structured as follows. Section \ref{sec:ABC} poses the problem of substructure lensing in a Bayesian framework and reviews the basics of Approximate Bayesian Computing. In section \ref{sec:sim_setup}, we describe our parameterization for the subhalo mass function, our method for creating mock data sets, and the procedure to compute posterior distributions from the forward model. In Section \ref{sec:results}, we examine how the signal from different substructure models appears surfaces as flux ratio anomalies, and provide forecasts for constraints on the half-mode mass for under various levels of precision in image flux measurements. We use a cosmology with $\Omega_m=0.3$, $\Omega_{\Lambda}=0.7$ and $h=0.7$. We use the software package {\tt{lensmodel}} to solve the lens equation and fit smooth models to lensing observables \citep{KeetonGLens2003}.
	
	\section{Bayesian inference on the subhalo mass function}
	\label{sec:ABC}
	\begin{table*}
		\centering
		\caption{Definitions and descriptions for parameters relevant to Equation \ref{eqn:posterior3}}
		\label{tab:notation}
		\begin{tabular}{lccr} 
			\hline
			parameter & definition & description\\
			\hline
			$\bf{d}_n$ & data from the $n$th lens & positions, time delays, flux ratios \\ \\ 
			$\qsub$ & vector of hyper-parameters  describing & ($A_0,m_{\rm{hm}}$), spatial distribution \\ & the global subhalo mass function \\ \\ 
			$A_0$ & normalization of subhalo mass function & $1\%$ substructure mass fraction at $R_{\rm{Ein}} \approx A_0 = 2 \times 10^8 \msun^{-1}$ \\ & & mass range $10^6 \leq M_{200} \leq 10^{10} \left[M_{\odot} \right]$ \\ \\
			$m_{\rm{hm}}$ & half-mode mass & number of subhalos below $m_{\rm{hm}}$ is strongly suppressed\\ \\
			$\msub$ & defines parameters for an individual & subhalo positions, masses, density profiles\\ & substructure realization & & \\ \\ 
			$\qmacs$ & maximum-likelihood macromodel for $n$th lens& fits $n$th positions, time delays in presence of substructure\\ \\
			$\fmods$ & flux ratios computed in the forward model for $n$th lens& computed with $\qmacs$ as opposed to\\ & & true mass distribution \\ \\
			\hline		
		\end{tabular}
	\end{table*}
	
	In this section we describe how we infer the parameters describing subhalo populations within a Bayesian framework, and propose an Approximate Bayesian Computing (ABC) algorithm to contend with the highly stochastic nature of substructure lensing. Section \ref{ssect:overview} derives the expression for the posterior distribution of dark matter model parameters given a set of lensing observables. In Section \ref{ssect:ABCoverview}, we briefly review the technique of Approximate Bayesian Computing.
	
	\subsection{Connecting dark matter model parameters to lensing observables}
	\label{ssect:overview}
	
	The observables from a strong lens system are image time delays $\bf{t}$, positions $\bf{x}$, and fluxes $\bf{f}$.\footnote{To impose constraints, we actually use flux ratios in order to divide out the unknown source flux.} The data vector for the $n$th lens can be written $\data = \left\lbrace \bf{x}_n,\bf{t}_n,\fobs\right\rbrace$, and we represent the dataset for a sample of $N$ lenses as set $\Data = \left\lbrace\bf{d}_1,\bf{d}_2,...,\bf{d}_N\right\rbrace$. 
	
	Given a dark matter model with global properties described by a set of hyper-parameters $\qsub$, the desired posterior distribution for the parameters $\qsub$ is given by
	\begin{equation}
	\label{eqn:posterior0}
	p\left(\qsub | \Data \right) \propto \mathcal{L}\left(\Data \ | \ \qsub \right) \pi\left(\qsub\right)
	\end{equation}
	where $\pi\left(\qsub\right)$ is the prior probability for the parameters. In practice, $\qsub$ describes the shape and normalization of the subhalo mass function, the spatial distribution of subhalos, the density profile of subhalos, etc. 
	
	Since the data from each lens is independent, the joint likelihood in Equation \ref{eqn:posterior0} can be written as a product of the likelihoods for each lens
	\begin{equation}
	\label{eqn:joint_like}
	\mathcal{L}\left(\Data \ | \ \qsub \right) = \prod_{n=1}^{N} \mathcal{L} \left(\data \ | \qsub\right).
	\end{equation}
	
	We specify the model for each lens as a combination of two mass components. The first is a macromodel, which accounts for most of the mass of the deflector and its environment. For the $n$th lens, we denote this component $\qmacn$. The second component is the substructure population described by by $\qsub$. With these definitions, the components of $\qmacn$ are nuisance parameters which are marginalized out of the posterior
	\begin{equation}
	\mathcal{L}\left(\data \ | \ \qsub \right) \propto \int p\left(\data \ | \ \qmacn,\qsub \right) \pi\left(\qmacn\right) d \qmacn,
	\end{equation}
	where we have assumed that the macromodel $\qmacn$ is independent of the dark matter parameters $\qsub$, and introduce the prior $\pi\left(\qmacn\right)$. The assumption that $\qmac$ and $\qsub$ are independent is not formally correct, as parameters such as the Einstein radius may be informative of the total halo mass and the normalization of subhalo mass function. Working with real datasets, this information would need to be incorporated in the analysis. For the purpose of forecasting the possible constraints on $\qsub$, we examine the case of two fixed normalizations that span the expected range of substructure abundance for the halo masses implied by the distribution of Einstein radii in our mock data (see Section \ref{sec:sim_setup}). As we will demonstrate in Section \ref{sec:results}, the information content of each lens scales with the overall normalization, such that the bounds on a sample of lenses with diverse halos masses will be bound by the two limiting cases of the overall normalization we analyze.
	
	Dark matter models do not directly map $\qsub$ to a set of lensing observables. Rather, $\qsub$ specifies statistical distributions for the masses, positions, density profiles, etc. of the subhalos. Defining a vector $\msub$ that specifies a specific substructure realization, the likelihood becomes 
	
	\begin{eqnarray}
	\label{eqn:marginal_like}
	\nonumber \mathcal{L}\left(\data \ | \ \qsub \right) \propto \int \mathcal{L}\left(\data \ | \ \msub, \qmacn \right) p\left(\msub | \qsub\right) \\ \times \pi\left(\qmacn\right) d \msub d \qmacn
	\end{eqnarray}
	
	To make progress in evaluating Equation \ref{eqn:marginal_like}, we note that the astrometric and time delay perturbations from substructure are generally small and can be partially absorbed by small adjustments in $\qmacn$.\footnote{Positions and time delays have some ability to probe substructure \citep{Che++07,K+M09}, but flux ratios experience stronger perturbations that are our focus here.} The flux ratios, on the other hand, are determined by the second derivative of the gravitational potential near an image, thus the effects of substructure are difficult to reproduce by adjustments in $\qmacn$. With this in mind, we write image positions and time delays separately from the flux ratios, writing $\data = \left\lbrace \dlin, \fobs\right\rbrace$, where $\dlin$ denotes the positions and time delays $\left\lbrace\bf{t},\bf{x}\right\rbrace$. To relate $\qsub$ to $\dlin$ and $\fobs$, our strategy is to forward model simulated data sets of image positions, time delays, and fluxes $\left\lbrace\dlinmod\left(\qmacn,\msub\right),\fmod\left(\qmacn,\msub\right)\right\rbrace$, which depend on $\qsub$ through the realizations of substructure $\msub$. The likelihood of observing $\data$ is therefore
	
	\begin{equation}
	\label{eqn:forward_like}
	\mathcal{L}\left(\data \ | \ \msub ,\qmacn \right) = \mathcal{L}\left(\dlin \ | \dlinmod \right) \mathcal{L}\left(\fobs \ | \fmod \right).
	\end{equation} 
	Next, we note that most choices of $\qmacn$, with a wide prior distribution, yield the incorrect positions and time delays, and therefore do not contribute substantially to the integral in Equation \ref{eqn:marginal_like}. We therefore approximate the marginalization over the macromodel parameters by fixing the macromodel in a certain configuration $\qmacs$, which fits the image positions and time delays. This step avoids sampling the potentially vast parameter space of $\qmac$. Explicitly, $\qmacs$ is defined by the relation
	\begin{equation}
	\label{eqn:def_mlm}
	\dlin = \dlinmod\left(\qmacs,\msub\right).
	\end{equation}
	The procedure of re-optimizing the macromodel was also employed in \citet{D+K02}.
	
	By evaluating the flux ratios only with respect to $\qmacs$, we effectively take a derivative, while formally an integral is required to marginalize over $\qmac$. The procedure of optimizing, rather than marginalizing, the macromodel will yield a good approximation to the true likelihood as long as the image fluxes do not vary significantly over the range of macromodel parameters space for which the image positions and time delays are fit. We verify that the variation in image fluxes for different macromodel configurations that fit the observed image positions is smaller that the typical $5-8\%$ variations derived empirically in \citep{Gilman++17} by fitting smooth lens models to realistic deflectors. \footnote{We find that constraining the magnitude of the external shear, external shear angle, axis ratio and position angle to $0.01, 5^{\circ}, 0.05, 5^{\circ}$, as is possible with detailed modeling \citep[e.g.][]{Wong++17}, is sufficient to ensure the flux variations associated with marginalizing the macromodel is below $4\%$ per image. We also verify that uncertainty associated with the power-law slope of the main deflector does not incur a serious bias in our forecasts (see Appendix \ref{app:E}).} In macromodel parameterizations more complicated than power-law ellipsoids that possess additional degrees of freedom, the fluxes may vary substantially even for configurations of $\qmac$ with fixed image positions, and the macromodel may be better able to absorb flux perturbations from substructure. Stellar disks fall into this category \citep{Hsueh++16,Gilman++17}, as do models with extreme angular structures \citep{C+K05}, but the former are unlikely to be present in a sample of massive elliptical deflectors, and the latter are unphysical. If external sources of error were reduced such that the dominant source of flux uncertainty stemmed from marginalizing the macromodel, one would need to explicitly do the marginalization. The procedure outlined here should be amended to sample prior distributions of $\qsub$ and $\qmac$ constructed on a lens-by-lens basis when working with real data, but for the purpose of computational expediency and making approximate forecast statements we leave this level of detail for future work.
	
	After replacing the integral over $\qmac$ with $\qmacs$, Equation \ref{eqn:marginal_like} becomes
	\begin{equation}
	\label{eqn:posterior3}
	\mathcal{L}\left(\data | \qsub \right) \propto \int \mathcal{L}\left(\fobs \ | \fmods \right) p\left(\msub,\qsub \right) d\msub 
	\end{equation}
	where we introduce the notation $\fmods = \fmod\left(\qmacs, \msub\right)$. The parameters relevant to Equation \ref{eqn:posterior3} are summarized in Table \ref{tab:notation}. 
	
	At this step, a Markov Chain Monte Carlo integration scheme would be inefficient, as the flux ratios corresponding to the overwhelming majority of realizations $\msub$ would not match those observed in the data. Rather than computing Equation \ref{eqn:posterior3} directly, we employ a computational method that allows us to efficiently explore the parameter space spanned by $\qsub$. 
	
	\subsection{Approximate Bayesian Computing}
	\label{ssect:ABCoverview}
	Approximate Bayesian Computing (ABC) is a computational algorithm rooted in Bayesian statistics that circumvents the direct calculation of intractable likelihoods, and enables inferences from simulated data sets computed in a forward model. For details in addition to those presented in this section, see e.g. \citet{TurnerZandtABC,Csillery++10,Lintusaari++17}. In recent years, ABC has seen applications across a wide range of problems in cosmology and astrophysics \citep{Weyant++13,Robin++14,Hahn++17,Birrer++17,Herbel++17}. 
	
	In an implementation of ABC, one draws samples from a prior probability distribution, creates a forward model of simulated data from the samples, compresses the data sets into summary statistics (optional, but often necessary to keep computation costs low), and accepts or rejects the samples based on the similarity of the simulated to the observed data. An implementation of the algorithm therefore proceeds as follows:
	
	\begin{itemize}
		\item[\bf 1.] Sample from a set of model parameters $\bf{\theta}$.
		\item[\bf 2.] From the samples $\bf{\theta}$, forward-model a set of simulated data $\bf{d^{\prime}}$.
		\item[\bf 3.] The data vector $\bf{d^{\prime}}$ is often multi-dimensional, but in many cases the relevant information that will discriminate between different parameters $\bf{\theta}$ is contained in only a subset of the data. To reduce the dimensions of the problem, introduce the summary statistics $S(\bf{d^{\prime}})$ and $S(\bf{d})$, which compress the relevant information contained in $\bf{d^{\prime}}$ and $\bf{d}$.
		\item[\bf 4.] Introduce a distance metric $R\left(S(\bf{d}),S(\bf{d}^{\prime})\right)$ - for instance, the Euclidean distance between the summary statistics in N-dimensions - and accept the proposition $\bf{\theta}$ under the requirement
		
		\begin{equation}
		\label{eqn:distance_met}
		R\left(S({\bf{d^{\prime}}}),S(\bf{d})\right) \leq \epsilon
		\end{equation}
		for some tolerance $\epsilon$. 
		\item[\bf 5.] Repeat steps [1-4] until the distribution of accepted samples is stable to changes in the total number of samples computed, and the total number of samples retained in the posterior.
		
	\end{itemize}
	
	Formally, when implementing ABC one obtains samples from the posterior density
	
	\begin{equation} \label{eqn:ABCpos}
	p\left(\bf{\theta} | R\left(\bf{d^{\prime}},\bf{d}\right) \leq \epsilon \right),
	\end{equation}
	with the property
	\begin{equation}
	p\left(\bf{\theta} | \bf{d} \right) = \lim_{\epsilon\to\ 0} \left[ p\left(\bf{\theta} | R\left(\bf{d^{\prime}},\bf{d}\right) \leq \epsilon \right)\right].
	\end{equation} 
	Thus, assuming the summary statistic contains the information necessary to distinguish between different models, the distribution of accepted samples from $\bf{\theta}$ converges to the true posterior as $\epsilon$ tends to zero. Put another way, the relative number of accepted samples between multiple competing models reflects the relative probabilities of these models as $\epsilon \to 0$. In practice, one must compromise between an $\epsilon$ large enough to ensure timely convergence of the ABC procedure, and a value stringent enough to ensure the distribution of accepted samples is representative of the true posterior. 
	
	Crucial steps in the implementation of ABC include the choice of summary statistic $S(\bf{d})$, and the acceptance criterion $\epsilon$. A summary statistic which erodes the discriminating information contained in the data will not converge to the true posterior. In a similar vein, an acceptance threshold too lax will result in a posterior distribution too broad, with the extreme limit of accepting all samples from $\bf{\theta}$ and returning the prior. For this reason, assuming the algorithm has converged, the joint posterior distribution for the model parameters approximated in ABC will always be conservative, in the sense that it contains more volume than the true posterior. 
	
	In the context of substructure lensing, we compute a summary statistic for each realization based on the observed flux ratios, or the fluxes of three images normalized by the flux of the fourth. The summary statistic we use based on the observed flux ratios $\fobs$ and the forward model flux ratios $\fmods$ is given by 
	\begin{equation}
	\label{eqn:summary_stat}
	S \left(\fobs,\fmods \right) = \sqrt{\sum_{i=1}^{3} \left(\fobsi - \fmodsi\right)^2},
	\end{equation}
	where the summation runs over the three flux ratios of the $n$th lens, making use of the full information contained in these data. For an example of results using a different summary statistic, see Appendix \ref{app:C}.
	
	\begin{figure}
		\includegraphics[clip,trim=0cm 0cm 0cm
		0cm,width=.48\textwidth,keepaspectratio]{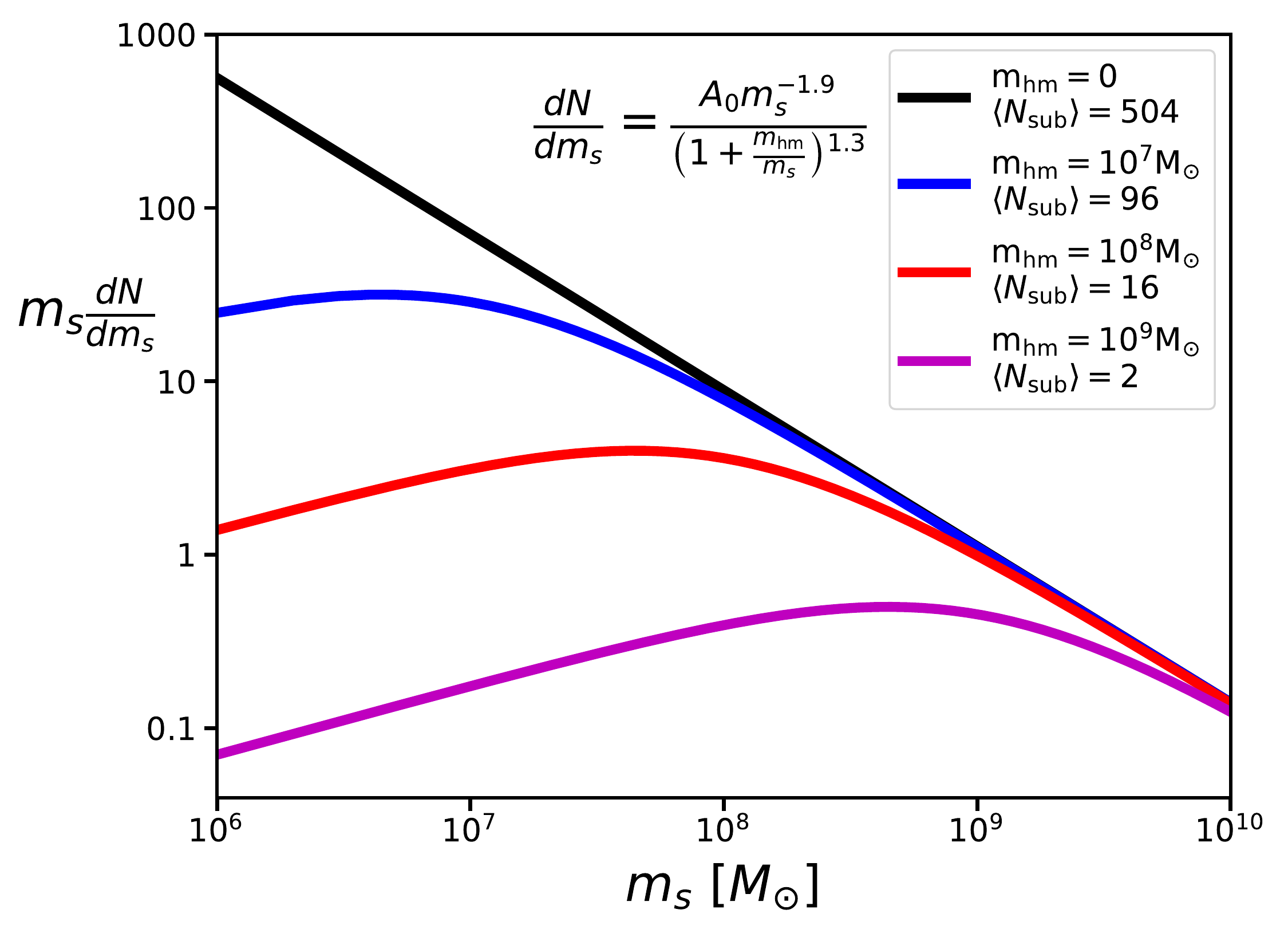}
		\caption{\label{fig:mass_function} The subhalo mass function of Equation \ref{eqn:wdm_massfunc} which we use to generate substructure realizations. In the figure we vary the half-mode mass $m_{\rm{hm}}$ with fixed normalization $A_0 = 1.2 \times 10^8 \msun^{-1}$. We generate subhalo populations in such a way that the amplitude of a CDM-like and a WDM-like mass functions are identical for masses $m \gg m_{\rm{hm}}$, rendering subhalos in projection to a radius of 18.6 kpc.}
	\end{figure}
	
	\section{Simulation Setup}
	\label{sec:sim_setup}
	In this section, we describe our lensing simulations, including the ingredients necessary to render substructure realizations and how we implement ABC to constrain the subhalo mass function. Section \ref{ssec:substructure} describes our prescription for modeling substructure populations for both cold and warm dark matter scenarios. Section \ref{ssec:mockdata} describes the mock data sets we use in our simulations, including lens and source redshift configurations. In section \ref{ssec:to_posterior}, we explain how we use the information contained in the forward model within the ABC framework to make inferences on $\qsub$. 
	
	\subsection{Subhalo density profile, mass definition, and mass function}
	\label{ssec:substructure}
	When quoting subhalo masses, we refer to the mass inside a sphere of radius $r_{200}$, $M_{200}$. We model subhalos as tidally truncated NFW profiles \citep{Baltz++09}
	\begin{equation}
	\label{eqn:massprofile}
	\rho \left(r\right) = \frac{\rho_0}{x \left(1+x\right)^2} \frac{\tau^2}{x^2 + \tau^2}
	\end{equation}
	where $x = \frac{r}{r_s}$, $\tau = \frac{r_t}{r_s}$, $r_t$ is the truncation radius and $r_s$ is a scale radius. \footnote{While experimenting with Pseudo-Jaffe profiles, we find that our results depend sensitively on how convergence is partitioned between the truncation radius and the central density. Since the central density dominates the flux ratio signal, different choices for the truncation and central density normalization yield significantly different results.} The finite, truncated mass can differ from $M_{200}$ to a varying degree depending on the truncation radius and the concentration, but the effect on image flux ratios is primarily determined by the central density, with the truncation playing a sub-dominant role provided $r_t > r_s$. 
	
	We render subhalos with $M_{200}$ (denoted $m_s$) between $10^6 \leq \frac{m_s}{M_{\odot}} \leq 10^{10}$, drawing from a subhalo mass function written as a broken power law, 
	\begin{eqnarray}
	\label{eqn:wdm_massfunc}
	\nonumber \frac{dN_{\rm{wdm}}}{dm_s} & = & \frac{dN_{\rm{cdm}}}{dm_s} \left(1+\frac{m_{\rm{hm}}}{m_{\rm{s}}}\right)^{-1.3} \\
	& = & A_0 \left(\frac{m_{\rm{s}}}{\msun}\right)^{-1.9} \left(1+\frac{m_{\rm{hm}}}{m_{\rm{s}}}\right)^{-1.3},
	\end{eqnarray}
	a functional form which resembles the subhalo mass function of a WDM particle \citep{Schneider++12,Lovell++14}. We restrict our analysis to this specific parameterization, and do not attempt to constrain the exponent -1.3 appearing in Equation \ref{eqn:wdm_massfunc} or the -1.9 slope of the CDM function. More complex scenarios, such as multi-component dark matter in which only a fraction of the dark matter is warm, will require a more careful treatment of the parameterization \citep[see][]{Vegetti++18}. For reference, $A_0 \approx 2 \times 10^8 \msun^{-1}$ corresponds to a convergence in substructure of 0.005, or a mass fraction of $1\%$ at the Einstein radius (for more details see Appendix \ref{app:A}). 
	
	The parameter $m_{\rm{hm}}$, the half-mode mass, denotes the mass scale at which the WDM power spectrum is damped with respect the CDM case by one-half. Assuming a thermal relic particle of mass $m$ comprises the dark matter, one can establish scaling relation $m_{\rm{hm}} \propto m^{-3.33}$ \citep{Schneider++12}. We normalize this relation to the $2\times 10^8 \msun h^{-1} \sim 3.3 \rm{keV}$ result of \citet{Viel13}, and translate between the two parameters as 
	\begin{eqnarray}
	\label{eqn:masskev}
	m_{\rm{hm}}\left(m\right) & = & 10^{10} \left(\frac{m}{\rm{1keV}}\right)^{-3.33} \msun h^{-1}.
	\end{eqnarray}
	
	In our simulations, we convert to physical masses in the lens plane using $h=0.7$. With this metric, the 2 keV result from \citet{Birrer++17} corresponds to $m_{\rm{hm}} = 10^{8.8} \msun$. As shown in Figure \ref{fig:mass_function}, we normalize the mass function such that $m_{\rm{hm}}$ does not affect the amplitude at scales $m_s \gg m_{hm}$, yielding the same numbers of very massive subhalos, on average.
	
	We combine this mass definition with a form for the mass-concentration relation for warm dark matter halos presented by \citet{BoseHellwing++16} \citep[see also][]{Maccio++08,Ludlow++16}
	\begin{equation}
	\label{eqn:cmrelation}
	c\left(m_s\right) = 6 \left(\frac{m_{\rm{s}}}{10^{12}\msun}\right)^{-0.098} \left(1+60\frac{m_{\rm{hm}}}{m_{\rm{s}}}\right)^{-0.17}.
	\end{equation} 
	which results in lower central densities at a given mass for warm dark matter models. The relation between concentration and $m_{\rm{hm}}$ reflects the later collapse epoch of small WDM subhalos, which prevents them from building up their concentrations over time. \footnote{We do not model the scatter in the mass-concentration relation. In a careful measurement, this feature should be included.}
	
	Given an $M_{200}$ drawn from the mass function in Equation \ref{eqn:wdm_massfunc}, and a concentration from Equation \ref{eqn:cmrelation}, we compute the normalization $\rho_0$ and the scale radius $r_s$. To obtain the truncation radius, and the lensing properties associated with the mass profile in Equation \ref{eqn:massprofile}, we generate subhalos in a 3-D sphere of radius 250 kpc (see Appendex \ref{app:A} for details regarding the spatial distribution). Given $r_{3d}$, we compute the truncation radius \citep{CyrRacine++16}
	
	\begin{equation} 
	\label{eqn:truncrad}
	r_t = \left(\frac{m_{s} r_{\rm{3d}}^2}{2 \Sigma_{\rm{crit}} R_{\rm{Ein}}}\right)^{\frac{1}{3}},
	\end{equation}
	where $\Sigma_{\rm{crit}}$ is the critical density and where $R_{\rm{Ein}} \approx 1''$ is a typical Einstein radius. 
	
	\subsection{Mock Data Sets}
	\label{ssec:mockdata}
	We consider three subhalo mass functions: a WDM mass function with $m_{\rm{hm}}$of $10^8 \msun$ and normalization of $10^8 \msun^{-1}$, and two CDM mass functions with normalizations of $8\times 10^7 \msun^{-1}$ and $40 \times 10^7 \msun^{-1}$. The two normalizations in the CDM simulations correspond to projected mass fractions at the Einstein radius of $0.4\%$ and $2\%$, respectively and bracket a plausible range than spans the theoretical uncertainties associated with the connection to halo mass \citep[e.g.][]{JiangBosch++17} and the tidal destruction of subhalos by the host galaxy \citep{DespVeg16,GK++17}. The low normalization case corresponds to the scenario in which all subhalos inside the halo NFW scale radius are destroyed, while the latter corresponds to no destruction (see the discussion in Appendix \ref{app:A} for details on obtaining these numbers).
	
	When rendering subhalos to create mock data sets, we solve the lens equation and ray trace with every subhalo between $10^6$ and $10^{10} \msun$ included in the computation. We distribute substructures over an SIE+shear macromodel with randomly oriented shear and ellipticity. Ellipticity (shear) is sampled from a Gaussian with mean 0.2 (0.05) and standard deviation 0.05 (0.01), and Einstein radii sampled from a Guassian with mean 1" and variance 0.2". We randomly sample source positions to produce equal numbers of cusp, fold and cross configurations. In Appendix \ref{app:B}, we 
	compare the sensitivities of the different image configurations by using each on separately to infer $\qsub$. We take the deflector to lie at a typical redshift $z_d = 0.5$, and put the source at $z_s = 1.5$. The source in both the data and forward model is parameterized by a circular Gaussian with width of 10 pc, or 1.2 m.a.s. To check to what degree this source size plays a role in our analysis, using source sizes as large as 30 pc we generate distributions of thousands of flux ratios from identical mass functions, and verify the distributions are nearly identical. 
	
	Even in the presence of identical subhalo populations, the observed and simulated flux ratios will differ due to the underlying macromodel and measurement errors. Examining mock lens systems with luminous mass components from galaxies in the Virgo and Coma clusters, in \citet{Gilman++17} we found flux ratio anomalies in mock deflectors composed of a NFW halo with a galaxy at its center, with respect to an SIE+shear model. Ray tracing through galaxies formed in the Illustris simulations, the authors of \citet{Hsueh2++17} also conclude that the incorrect macromodel can contribute to a measured flux ratio anomaly. 
	
	To ensure that modest deviations away from an isothermal-ellipsoid macromodel parameterization does not bias the precision of our inference on $\qsub$, we have simulated mock lenses with with power law slopes drawn from a distribution offset from isothermal and modeled them with SIE profiles in the forward model. Encouragingly, the precision of our forecast is not degraded when subjected to this source of error. See Appendix \ref{app:E} for more details regarding this test. In practice, systematic error associated with the macromodel can be dealt with by sampling additional parameters in the forward model. For instance, macromodel deviations around an isothermal profile can be handled by simultaneously sampling the power law slope and $\qsub$. \footnote{In principle, this approach could be extended in the forward modeling framework to more complex morphological features on a lens by lens basis.}
	
	Even for samples of morphologically simple deflectors (i.e. no disks or other prominent morphological features), we expect deviations in flux ratios at the percent level from measurement errors, and residual uncertainties in the image fluxes caused by the parameterization of the macromodel. To incorporate these uncertainties in our simulations, we add perturbations to the fluxes in our mock data sets. We model flux anomalies as Gaussian, and perturb each flux $F$ as
	\begin{eqnarray}
	\label{eqn:newfr}
	F & \rightarrow & F + \delta F; \\ \delta F & = & \nonumber \mathcal{N} \left(0,\delta \times F\right),
	\end{eqnarray}
	examining specific cases of $\delta = 0.02$ $\delta = 0.04$ and $\delta = 0.08$, which correspond to flux errors of $2,4$ and $8\%$. Conceptually, one can interpret this operation as erasing information at the $\delta$ level, which enables one to track the sensitivity of lensing constraints on small sources of flux perturbations. These perturbations lump together all deviations in image fluxes away from those of an SIE+shear model fit to the data that are not caused by dark substructure, including measurement errors and the baryonic structure of the deflector, and bracket the range of errors one excepts for morphologically simple deflectors. These perturbations are empirically motivated by the flux residuals we encounter in \citep{Gilman++17} fitting smooth lens models to lenses build from high resolution images of galaxies in the nearby Virgo cluster. We reiterate that extreme morphological features like stellar disks may produce larger systematic perturbations than those we are mimicking with the $\delta$ perturbations, but these prominent features are unlikely to be present in massive ellipticals \citep{Auger:2010p30340,Son++13a}. \footnote{In fact, studies of massive elliptical lensing galaxies find their mass profiles are well modeled by nearly isothermal power law ellipsoids \citep{SonnenfeldEtal2015,Shankar++17,Gilman++17}.} Furthermore, deflectors likely to contain disks can be readily identified based on velocity dispersion and stellar mass, in addition to high resolution imaging. Since we add these flux errors independent of perturbations to the image positions and time delays, we assume that the macromodel and measurement-error induced flux ratio anomalies are independent from astrometric and time delay anomalies, a conservative choice as correlations provide additional information that can be used to identify these features in the data.
	
	For reference, current techniques using measurements of narrow-line fluxes achieve precision of roughly $4-6\%$ \citep{Nie++14,Nierenberg++17}. The $8\%$ errors can therefore be interpreted as pessimistic case, with errors induced by the use of a simplified macromodel compounding measurement errors, while $2\%$ simulates an optimistic future precision. The $4\%$ curve serves to illustrate how the bounds evolve between these two extremes. Finally, we add measurement errors of 3 m.a.s. to the mock image positions, typical of astrometric uncertainties with current instruments.
	\begin{figure}
		\includegraphics[clip,trim=4.2cm 3.2cm 2.2cm
		2.7cm,width=.52\textwidth,keepaspectratio]{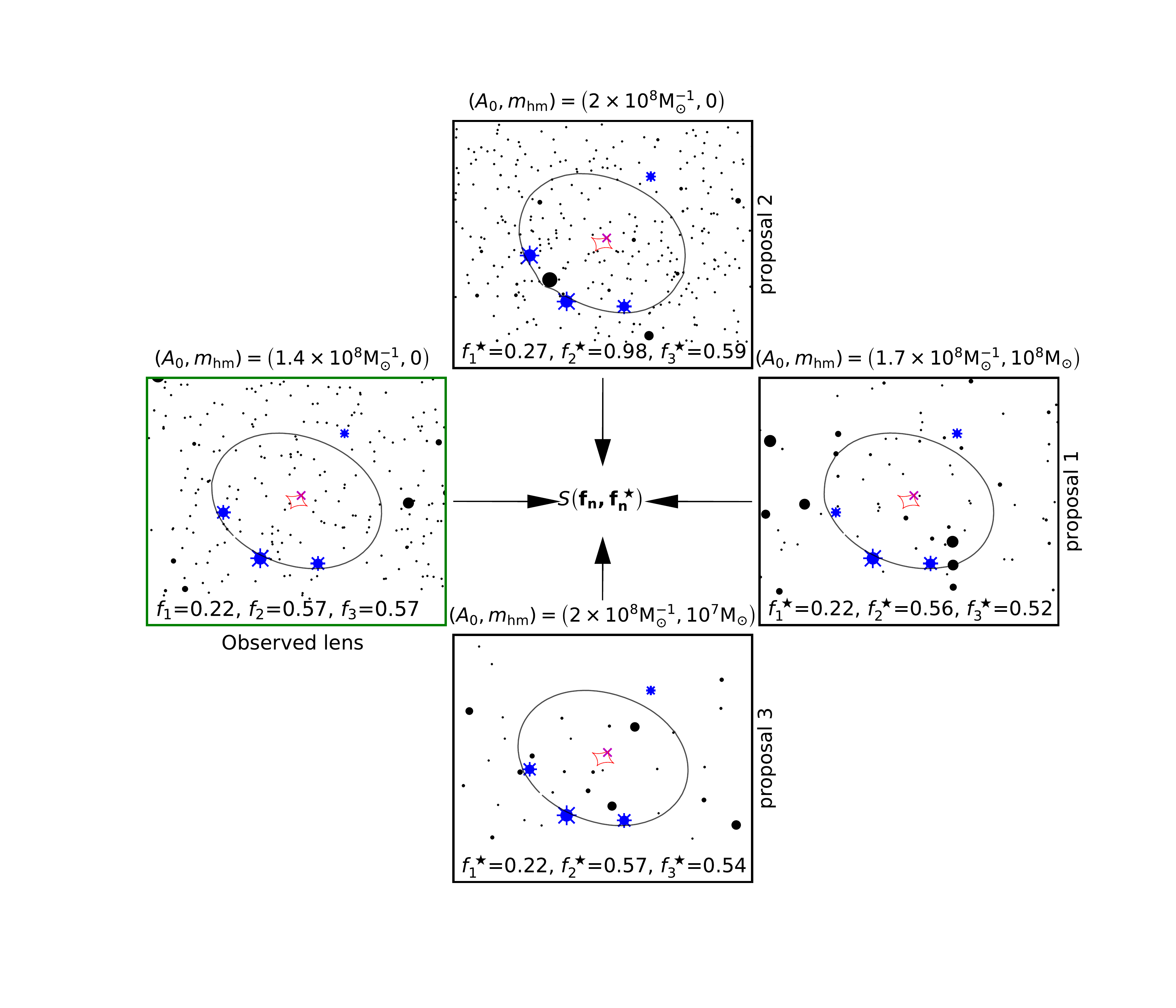}
		\caption{\label{fig:visualization2} This figure presents an illustration of the forward modeling analysis.  We use one realization of substructure to create mock data, as shown in the panel outlined in green (left). Lensed images (blue) have sizes corresponding to their flux, while subhalos (black) have sizes corresponding to $M_{200}$.  We compute flux ratios with respect to the middle image in the triplet; $f_1$ denotes the image in the upper right, while $f_2$ and $f_3$ denote images in the left and right of the triplet, respectively. We then discriminate between different parameters describing the subhalo mass function by drawing many substructure realizations from proposal mass functions, three examples of which are shown here (panels outlined in black). For each realization we re-optimize the macromodel and compute the summary statistic $S\left(\fobs,\fmods\right)$ from Equation \ref{eqn:summary_stat}, which we use to accept or reject the realization and the parameters describing the mass function it came from. The procedure visualized here is repeatedly applied in the full analysis shown in Figure \ref{fig:visualization}. 
		}
	\end{figure}
	\begin{figure}
		\includegraphics[clip,trim=4cm 5cm .5cm
		5cm,width=.45\textwidth,keepaspectratio]{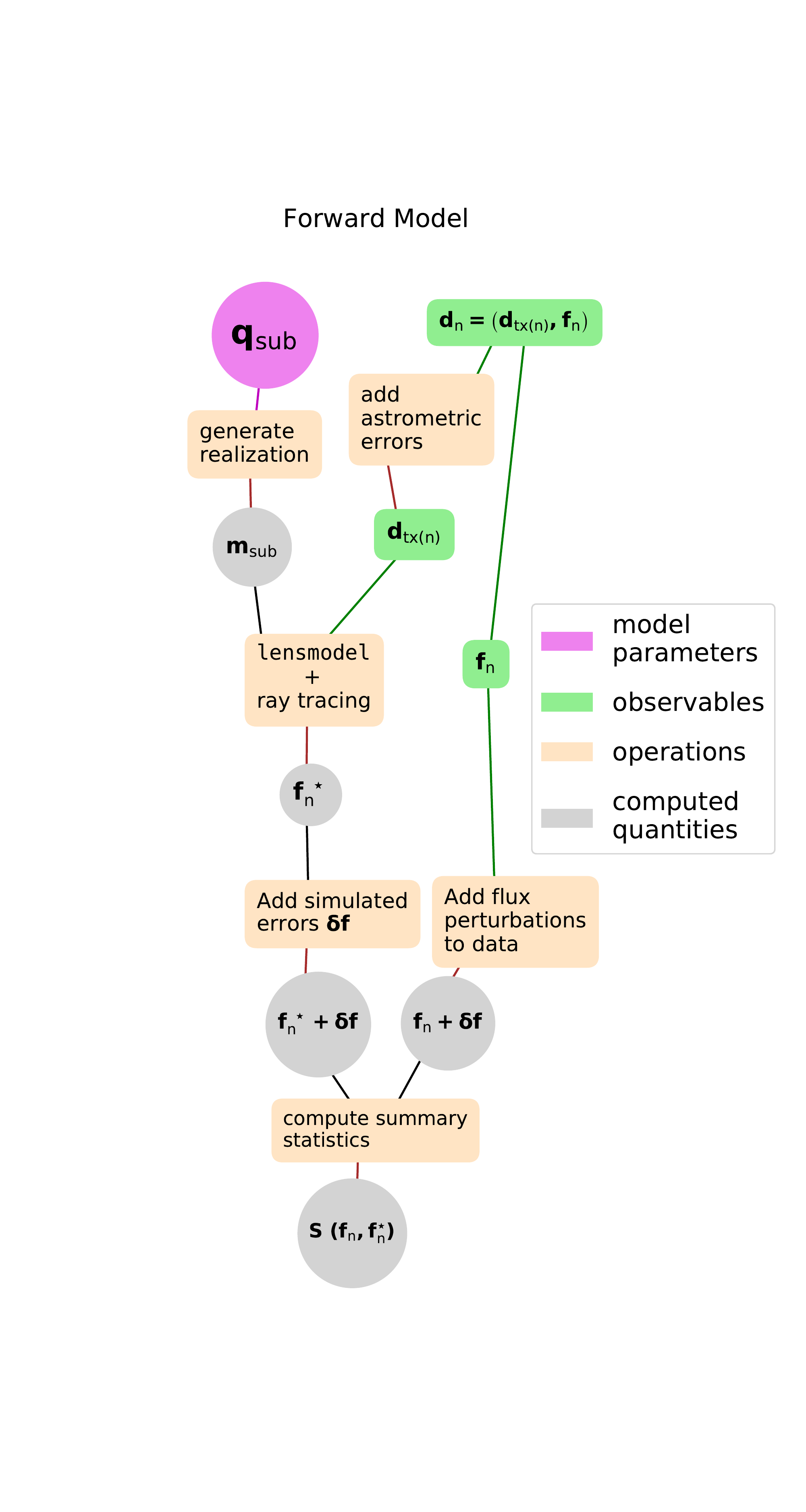}
		\caption{\label{fig:visualization} A schematic overview of the forward model used to compare flux ratios in the simulated data sets to flux ratios derived from substructure realizations drawn from $\qsub$.}  
	\end{figure}
	\subsection{Constraining the subhalo mass function}
	\label{ssec:to_posterior}
	In our simulations, the subhalo mass function is defined by the free parameters $A_0$ and $m_{\rm{hm}}$; our goal is to relate these parameters to the observed data from $N$ simulated lenses, $\bf{D}$. To do so, for each $A_0$ and $m_{\rm{hm}}$, we render $\approx 2000$ substructure realizations $\msub$ per proposed set of parameters $\qsub$, sampling the prior uniformly in $A_0$ and uniformly in $\log_{10}\left(m_{\rm{hm}}\right)$, yielding in $\approx 10^6$ realizations per lens. In Appendix \ref{app:D}, we perform tests to verify convergence with this number of samples. For each realization, we use {\tt{lensmodel}} to re-optimize the macromodel parameters $\qmacn$ that satisfy Equation \ref{eqn:def_mlm}, and ray-trace through a grid sampled at 0.4 m.a.s. per pixel to obtain image magnifications and flux ratios with an extended background source modeled as a circular Gaussian with a full-width at half-maximum of $10$ pc. When re-optimizing the macromodel, we assume uncertainties of 3 m.a.s. and 2 days on positions and time delays. We use flux constraints weak enough that they do not impact the re-optimization of the macromodel, allowing us to impose different flux perturbations in our mock data sets after running the simulation. 
	
	To deal with the flux errors we add to our mock data sets, we add perturbations of the same form as Equation \ref{eqn:newfr} to the forward model fluxes. With the perturbed fluxes in hand, we evaluate the statistic in Equation \ref{eqn:summary_stat}. After computing the summary statistics, each proposal of $\qsub$ in each of the $N$ lenses has a set metric distances associated with it. At this juncture, we apply a rejection criterion to select the most probable models, including the 1,800 samples with smallest corresponding summary statistics. In principle, because the systems are statistically independent, we could apply this criteria to each one individually and multiply the resulting distributions. Practically, however, multiplying a large number of probability densities together computed on a discretely sampled space is numerically unstable. To handle these numerical issues, we first reduce the dimensionality adding the summary statistics for pairs of lenses. In the limit of an infinite number of realizations and infinitely stringent acceptance criteria, this is equivalent to multiplying the likelihoods. To each of the resulting $\frac{N}{2}$ probability densities, we apply a Gaussian Kernel Density Estimator (KDE), and multiply them. We verify that, with these choices, our algorithm satisfies a rudimentary test of convergence (see Appendix \ref{app:D}).
	
	In summary: the forward model $\fmods$ contains the information needed to discriminate between different dark matter models through the realizations $\msub$, drawn from the hyper-parameters $\qsub$. For any parameterization of $\qsub$, Equation \ref{eqn:posterior3} relates the flux ratios $\fmod\left(\qmacs,\msub\right)$ to the observed data by evaluating the flux ratios at fixed image positions, enforced by first re-optimizing the macromodel by fitting a smooth lens model to the data $\dlin$, and then computing the flux ratios $\fmods$ for this lens model in the presence of the substructure realization $\msub$. 
	
	We account for flux ratio anomalies caused by measurement errors, and by the imperfect SIE+shear macromodel fit to realistic deflectors, by adding random Gaussian perturbations to the forward model fluxes. We then compute a summary statistic which reflects the degree to which the observed flux ratios match those computed in the forward model. The final posterior probability distribution is composed of the set samples from the prior $\pi\left(\qsub\right)$ whose corresponding realizations $\msub$ yield summary statistics closest to those computed from the flux ratios in the observed data. To keep computational costs low, in the forward model we only render substructures below masses of $10^{7.5} \msun$ if they lie within 0.5 arcseconds of an image. All higher mass subhalos are included regardless of position.
	
	\section{Results}
	\label{sec:results}
	
	Based on the method we outlined in the previous two sections, we are able to quantify the effect of substructure on image flux ratios, and forecast the constraining power of this method. We analyze three scenarios: a CDM mass function with a mass fraction in substructure of $0.4\%$ substructure at the Einstein radius $A_0 = 8 \times 10^7 M_{\odot}^{-1}$), a CDM mass function with a mass fraction in substructure of $2\%$ at the Einstein radius ($A_0 = 4 \times 10^8 M_{\odot}^{-1}$), and a WDM mass function with a mass fraction in substructure of $0.5\%$ substructure at the Einstein radius ($A_0 = 10^8 M_{\odot}^{-1}$) and a half-mode mass of $10^8 M_{\odot}$ corresponding to a 3.6 keV thermal relic. \footnote{Due to the demanding computational resources required per lens in the forward modeling procedure, we limit our present analysis to these cases.} We begin in Section \ref{ssec:signal} by discussing how variations in the normalization and half-mode mass impact distributions of flux ratios. In \ref{ssec:idealized_inf}, we show the results of simulations in which the mock data sets are free from errors; these simulations serve to determine the constraints achievable with the best possible data. In \ref{ssec:real_inf}, we add measurement errors and uncertainties to the fluxes in the mock data sets, account for them in the forward model, and quantify their effect on ones inference of $\qsub$. 
	
	\subsection{Flux ratio signal from structure in the lens plane}
	\label{ssec:signal}
	
	\begin{figure}
		\includegraphics[clip,trim=0cm 0cm 0.3cm
		0cm,width=.45\textwidth,keepaspectratio]{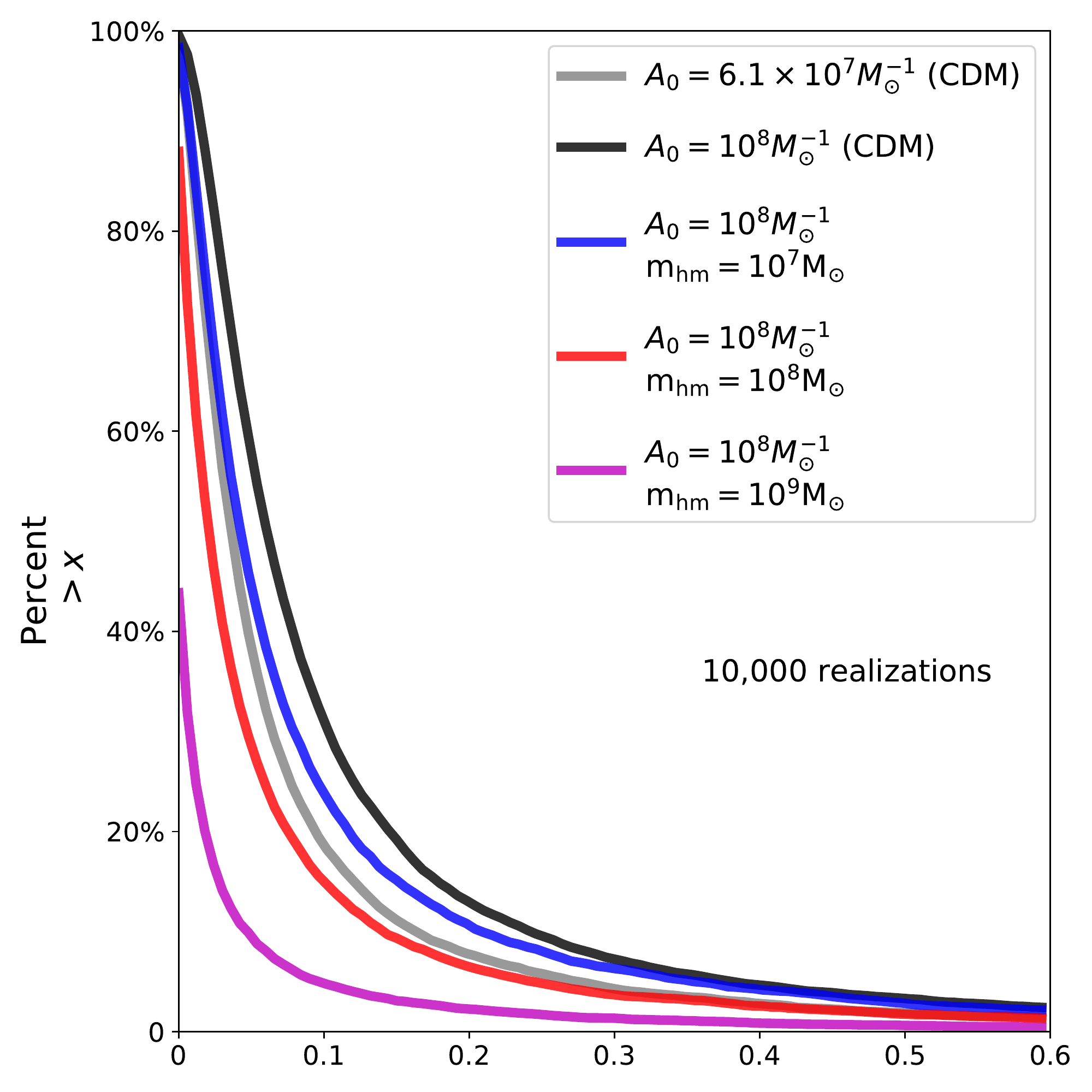}
		\includegraphics[clip,trim=0.2cm 0cm .3cm
		0cm,width=.45\textwidth]{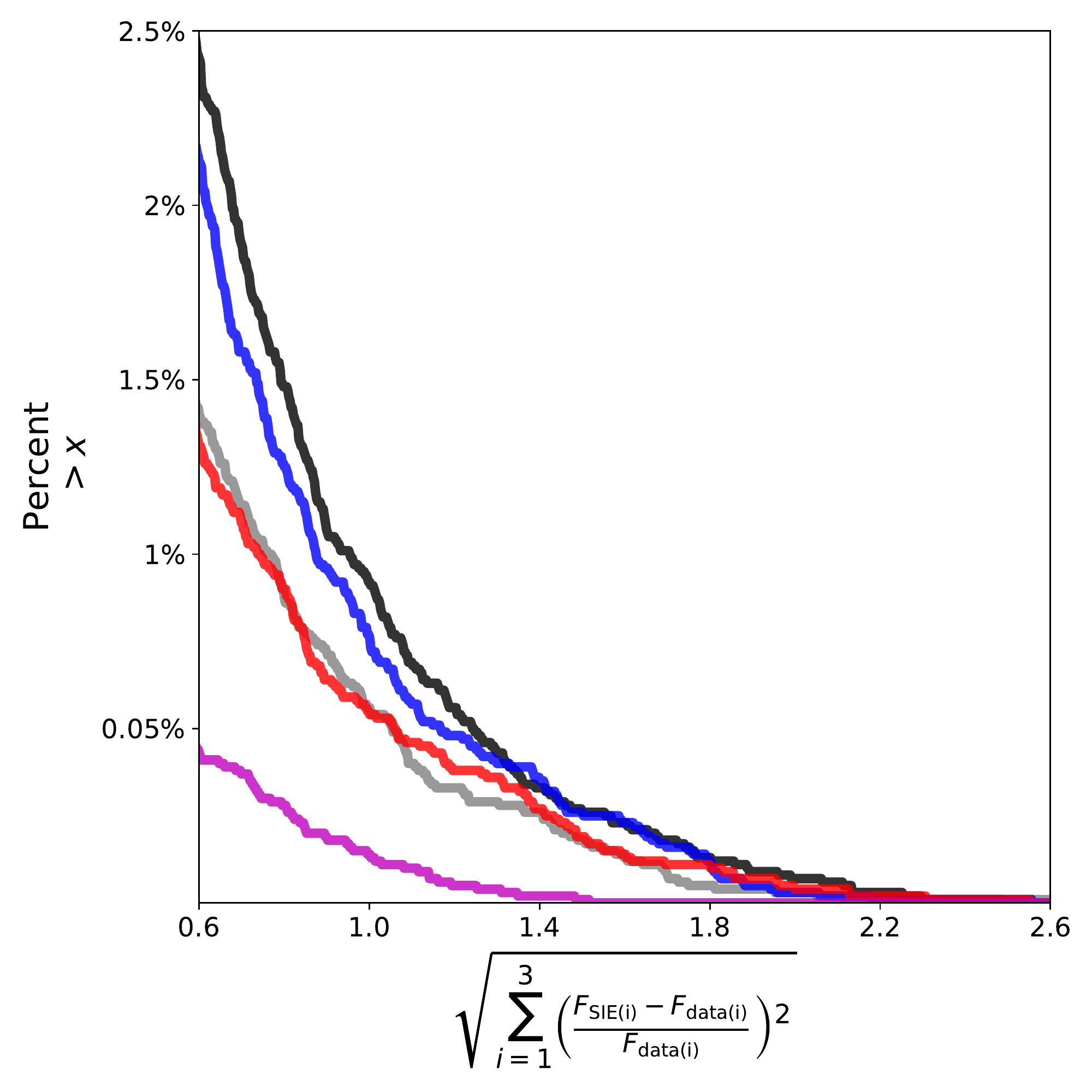}
		\caption{\label{fig:fluxratios} Cumulative distributions of flux ratio anomalies summed in quadrature for different subhalo mass functions with varying normalization (grey, black), and identical normalizations but varying $m_{\rm{hm}}$ (black, blue, red, magenta). The lower panel shows a zoom-in of the long, low probability tail of the distributions. Models with higher normalization (black vs. grey) produce more frequent flux ratio anomalies. Models with high $m_{\rm{hm}}$ produce less frequent anomalies than the black curve with $m_{\rm{hm}} = 0$ for flux ratio anomalies $<1.4$.}
	\end{figure}
	
	In Figure \ref{fig:fluxratios}, we plot the cumulative distribution of flux ratio anomalies for $10,000$ substructure realizations. Several trends emerge which help to understand the signal coming from substructure in image flux ratios.
	
	If we consider modest anomalies whose strength is $<60\%$ (summed in quadrature), we see that the normalization and half-mode mass both affect the frequency of anomalies. This suggests a degeneracy between the two mass function parameters, which indeed surfaces in a joint inference. The tails of the distributions, shown in the lower panel of Figure \ref{fig:fluxratios}, tell a different story: the curves behave similarly for anomalies whose strength is $>140\%$ (summed in quadrature), except for the most extreme WDM case (shown in magenta).  Together, these results suggest that the most massive substructures, which survive the free-streaming cutoff and are present in the black, blue, grey, and red curves, are responsible for the largest flux ratio anomalies. The frequency of flux ratio anomalies from the model with $m_{\rm{hm}}=10^9\msun$ suggests that substructures with masses between $10^7 \msun$ and $10^9 \msun$ dominate the lensing cross section for the largest flux ratio anomalies, while the cross section for small flux ratio anomalies is dominated by subhalos of mass $\lesssim 10^7 \msun$.
	
	\subsection{Inference on subhalo mass function with idealized data sets}
	\label{ssec:idealized_inf}
	
	Before adding simulated errors to the measured flux ratios, as will be present in a real sample of lenses, we first perform the inference on data sets where the flux ratios in the data and the forward model are un-perturbed. Effectively, in these simulations, the only unknowns are the properties of the underlying subhalo mass function, as the macromodel in both the data and the forward model is the same. They demonstrate the best one could hope to do by modeling subhalos only in the lens plane. 
	
	\begin{figure}
		\centerline{
			\includegraphics[clip,trim=0cm .5cm 0cm
			0cm,width=.48\textwidth,keepaspectratio]{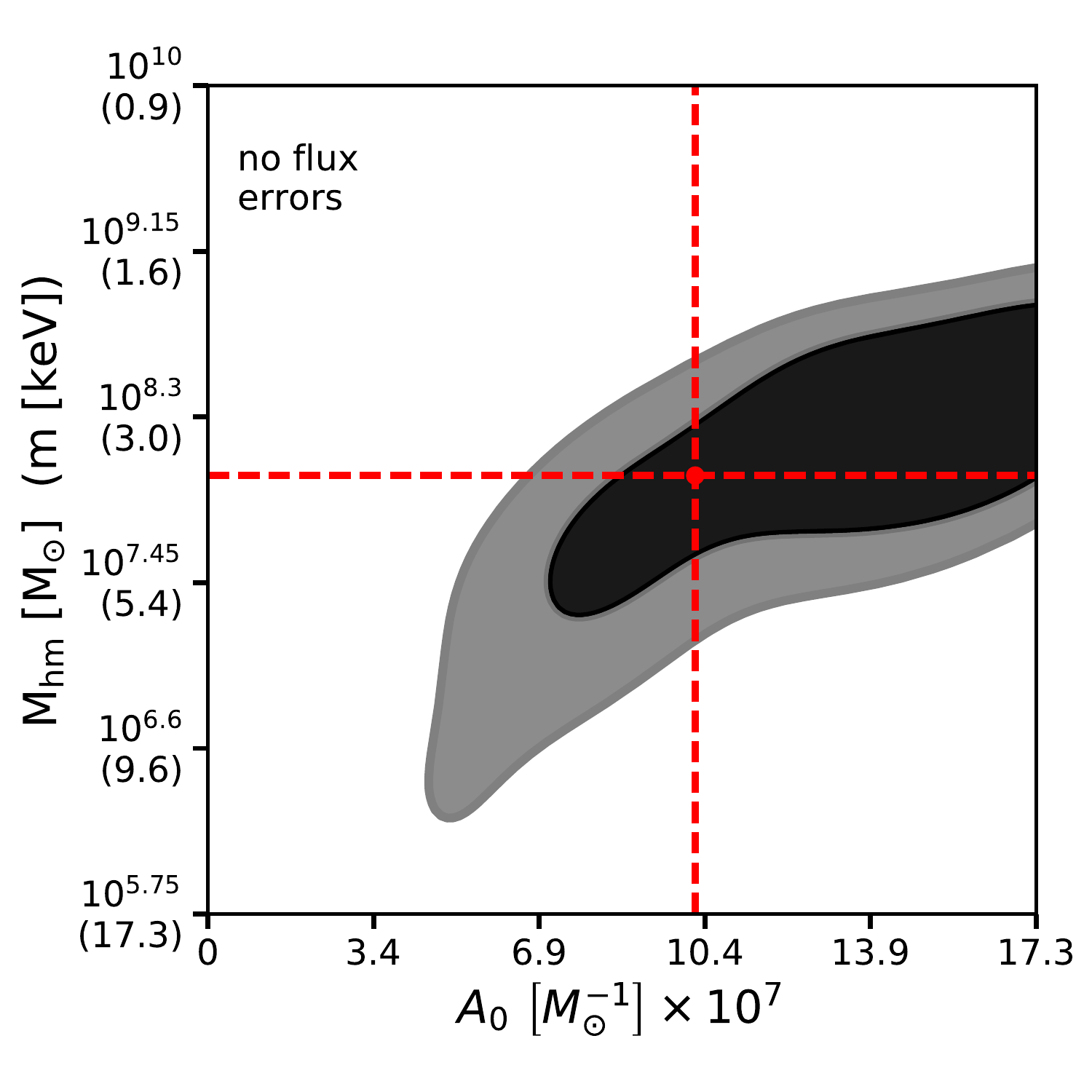}
		}
		\centerline{
			\includegraphics[clip,trim=0cm 0cm 0cm
			0cm,width=.45\textwidth,keepaspectratio]{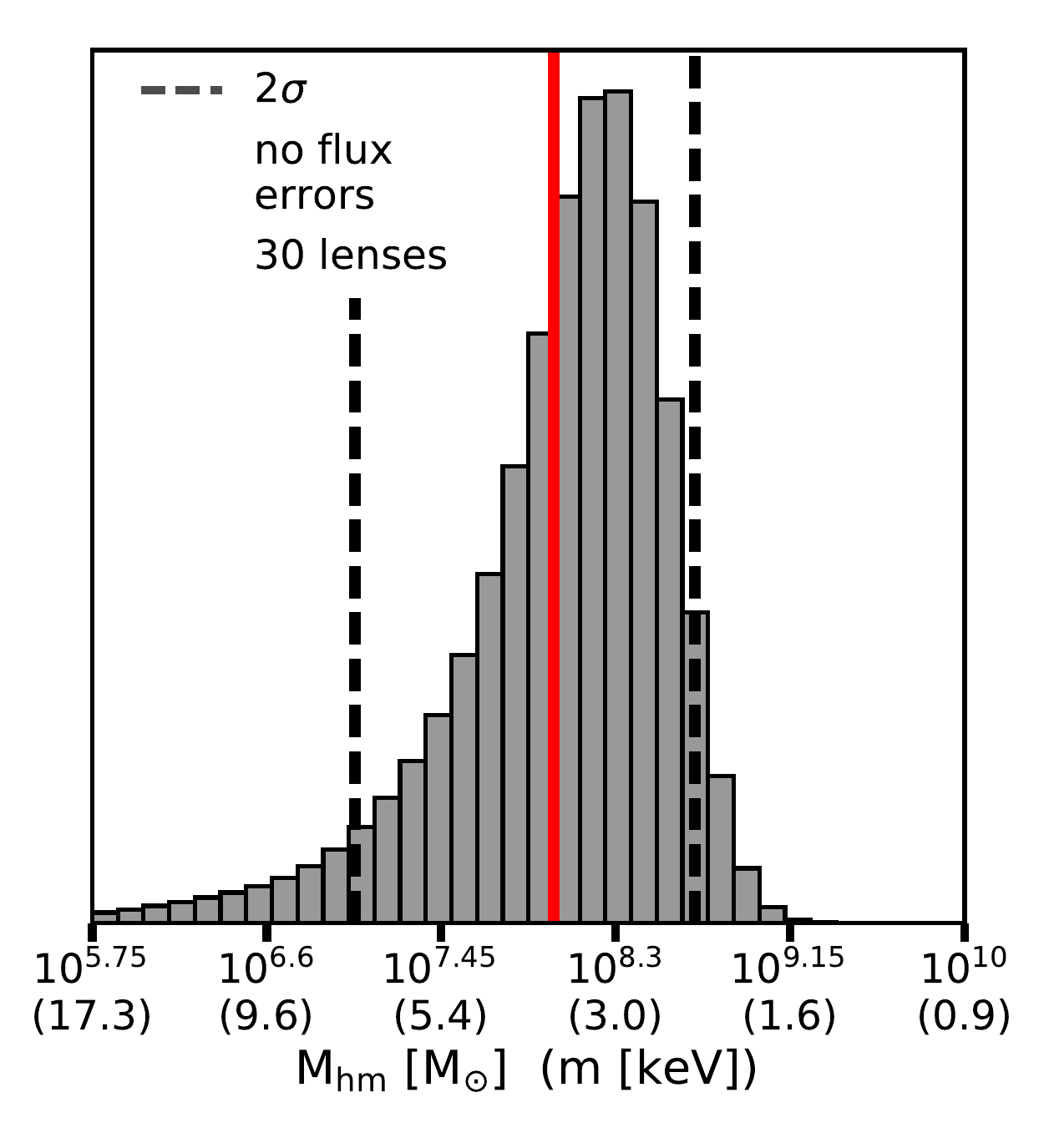}
		}
		\caption{\label{fig:wdm_inf_noerror}
			(Top) Joint posterior distribution for data with input $\left(A_0,m_{\rm{hm}}\right) = \left(1.04\times 10^8 \msun^{-1},10^8 \msun \right)$, marked by red lines. The closing of the $1\sigma$ contour from the bottom, ruling out CDM mass functions at $2\sigma$, demonstrates the sensitivity of the flux ratios, and this method, for probing substructure on scales $M_{200} \approx 10^{6.5} \msun$.
			(Bottom) Marginalized constraints on $m_{\rm{hm}}$.  The $2 \sigma$ bounds correspond to $10^{7} \msun < m_{\rm{hm}} < 10^{8.7} \msun$.}
	\end{figure}
	
	First, Figure \ref{fig:wdm_inf_noerror} shows the joint posterior distribution for a simulated data set of 30 lenses with $\qsub = \left(A_0,m_{\rm{hm}}\right) = \left(10.4 \times 10^7 \msun^{-1},10^8 \msun \right)$. This idealized calculation provides a useful limit to the sensitivity of the flux ratio anomaly method. The closing of the $2\sigma$ contour around $10^{6.5} \msun$ demonstrates that the signal in flux ratios is sensitive to subhalos of this mass. On their own, these objects produce a very weak lensing signal, but they create collective effects that should make it possible to distinguish between a CDM scenario in which they are abundant and a WDM scenario in which their numbers are depleted.  In thie case shown here, the $2 \sigma$ bounds on the half-mode mass are are $10^{7} \msun < m_{\rm{hm}} < 10^{8.7} \msun$, which correspond to bounds on the WDM particle mass of 7.3 and 2.3 keV, respectively. 
	
	\begin{figure}
		\includegraphics[clip,trim=0cm .5cm .3cm
		.5cm,width=.48\textwidth,keepaspectratio]{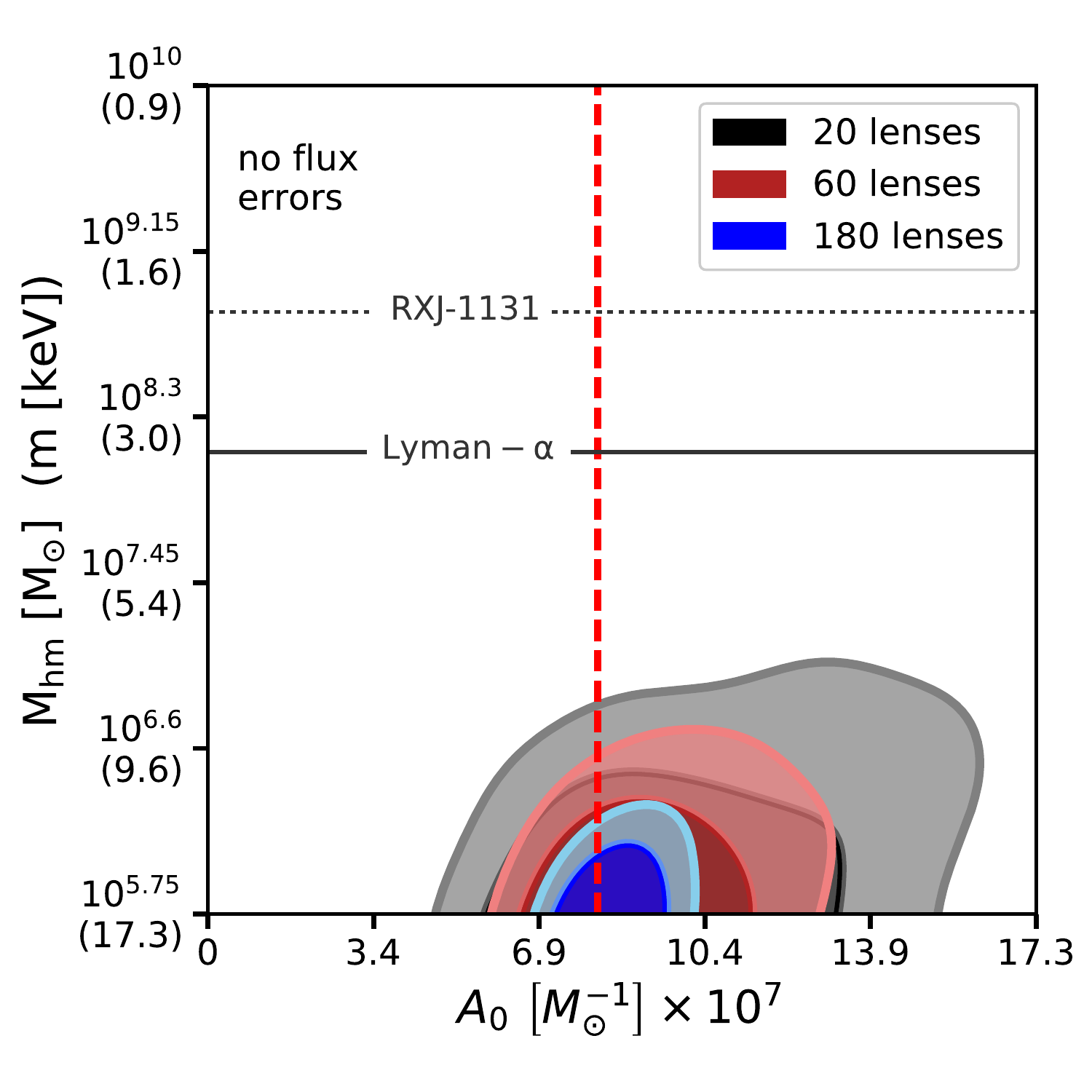}
		\caption{\label{fig:cdm_inf_noerror} Posterior distribution for a CDM-like mass function with input mass function parameters $\left(A_0,m_{\rm{hm}}\right) = \left(8.2 \times 10^7 \msun^{-1},0\right)$. In this idealized simulation, we do not add measurement errors or any other perturbations to fluxes in the simulated data sets. Effectively, the only unknown variables are $A_0$ and $m_{\rm{hm}}$, which describe the subhalo mass function. In such an idealized case, flux ratios probe scales below $10^{6.5} \msun$, ruling out WDM models with $m_{\rm{hm}} > 10^{6.1} \msun$ and $m_{\rm{hm}} > 10^{6.4}\msun$ at 1 and 2$\sigma$, respectively. For reference, grey solid and dashed lines show the $2\sigma$ bounds on a WDM particle mass from \citet{Viel13} and \citet{Birrer++17}, respectively.}
	\end{figure}
	
	In a second simulation, we use a data set composed of 180 systems with an input mass function with $\left(A_0,m_{\rm{hm}}\right) = \left(8.2\times 10^7 \msun^{-1},0 \right)$. Figure \ref{fig:cdm_inf_noerror} shows the joint posterior distribution on this data set as the number of lenses is increased. We interpret the $2 \sigma$ bound of $10^{6.4} \msun$ as the best one could do with 180 systems. \footnote{In each of the simulations with mock data sets containing CDM mass functions, we only quote the upper bound because the $2 \sigma$ lower bound is set by the limits of the prior assigned to $m_{\rm{hm}}$.} As we show in the next section, these bounds weaken significantly under flux ratio errors of $2\%$, $4\%$ and $8\%$, which mimic the signal in flux ratios produced by the smallest subhalos, or by subhalos far from an image.
	
	\subsection{Inference on subhalo mass function with simulated uncertainties}
	\label{ssec:real_inf}
	
	\begin{figure}
		\includegraphics[clip,trim=0cm .5cm .3cm
		.5cm,width=.48\textwidth,keepaspectratio]{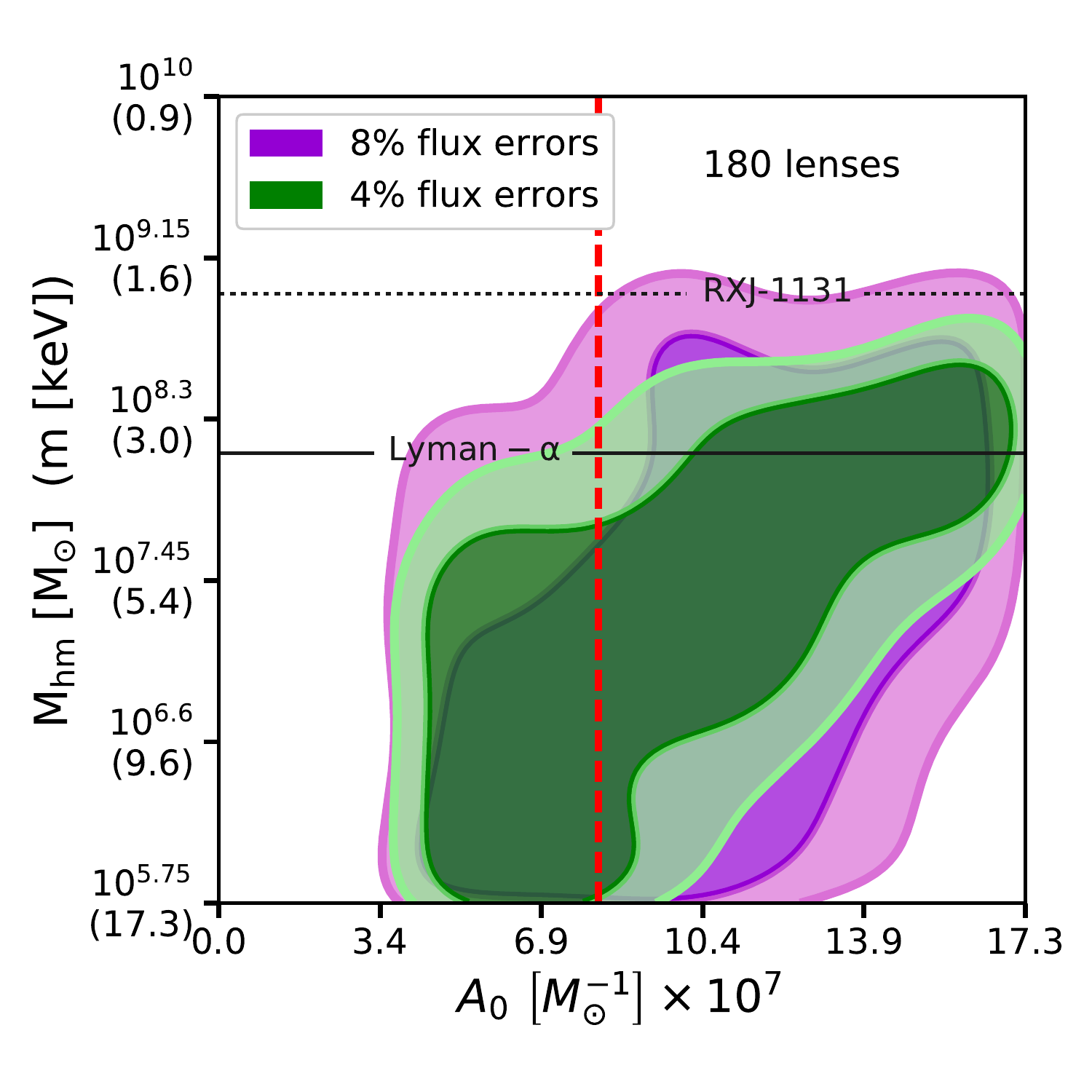}

		\caption{\label{fig:cdm_inf_error} Similar to Figure \ref{fig:cdm_inf_noerror}, but in this case the fluxes in the mock data sets receive $4\%$ and $8\%$ uncorrelated Gaussian errors. On their own, these errors look like substructure, biasing the inference to models with more subhalos. However, we are able to account for this uncertainty by introducing perturbations to the fluxes in the forward model that match those applied to the mock data. Adding noise to the fluxes washes out the signal from the smallest subhalos, and the $2\sigma$ constraining power on $m_{\rm{hm}}$ is diminished by over an order of magnitude. This figure shows the result of a single draw of Gaussian errors in the fluxes. In Figure \ref{fig:nlens_vs_2sigma}, we compute $2\sigma$ bounds on $m_{\rm{hm}}$ averaged over many draws of these errors.}
	\end{figure}
	
	\begin{figure}
		\includegraphics[clip,trim=0cm .5cm .3cm
		.5cm,width=.48\textwidth,keepaspectratio]{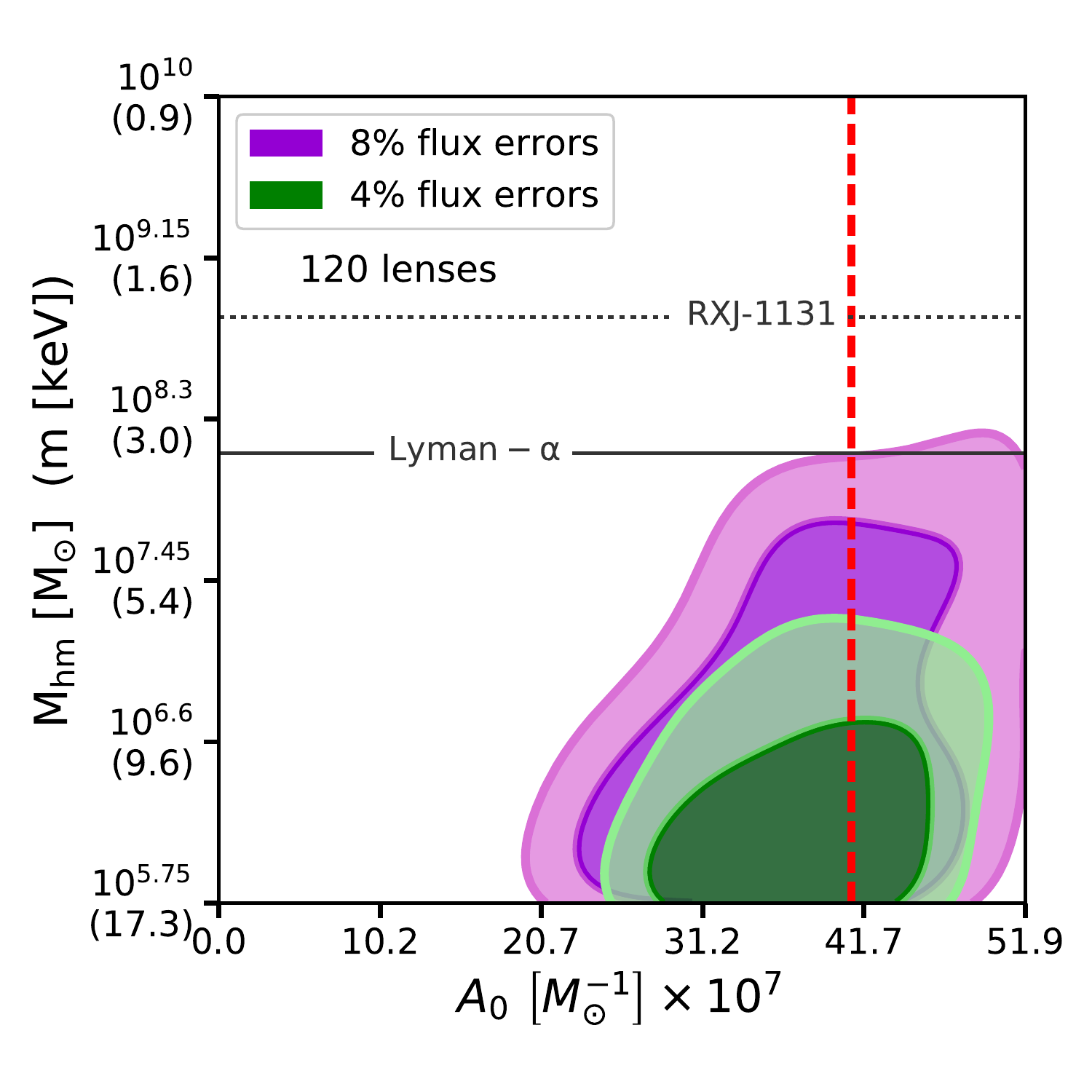}
		\caption{\label{fig:cdm_inf_error_highnorm} Similar to Figure \ref{fig:cdm_inf_error}, but an inference performed on a dataset with a subhalo mass function normalization corresponding to the expected mass fraction in substructure if there is no tidal disruption of subhalos within the host scale radius (see Appendix A). The higher abundance of substructures in this scenario results in a higher probability of observing flux perturbation larger than the effective detection threshold, which is determined by the precision of the image fluxes. Neither the high nor low normalization scenarios includes the expected boost from line of sight structure which we will consider in a future work.}
	\end{figure}
	
	\begin{figure}
		\includegraphics[clip,trim=0cm 0cm 0cm
		0cm,width=.48\textwidth,keepaspectratio]{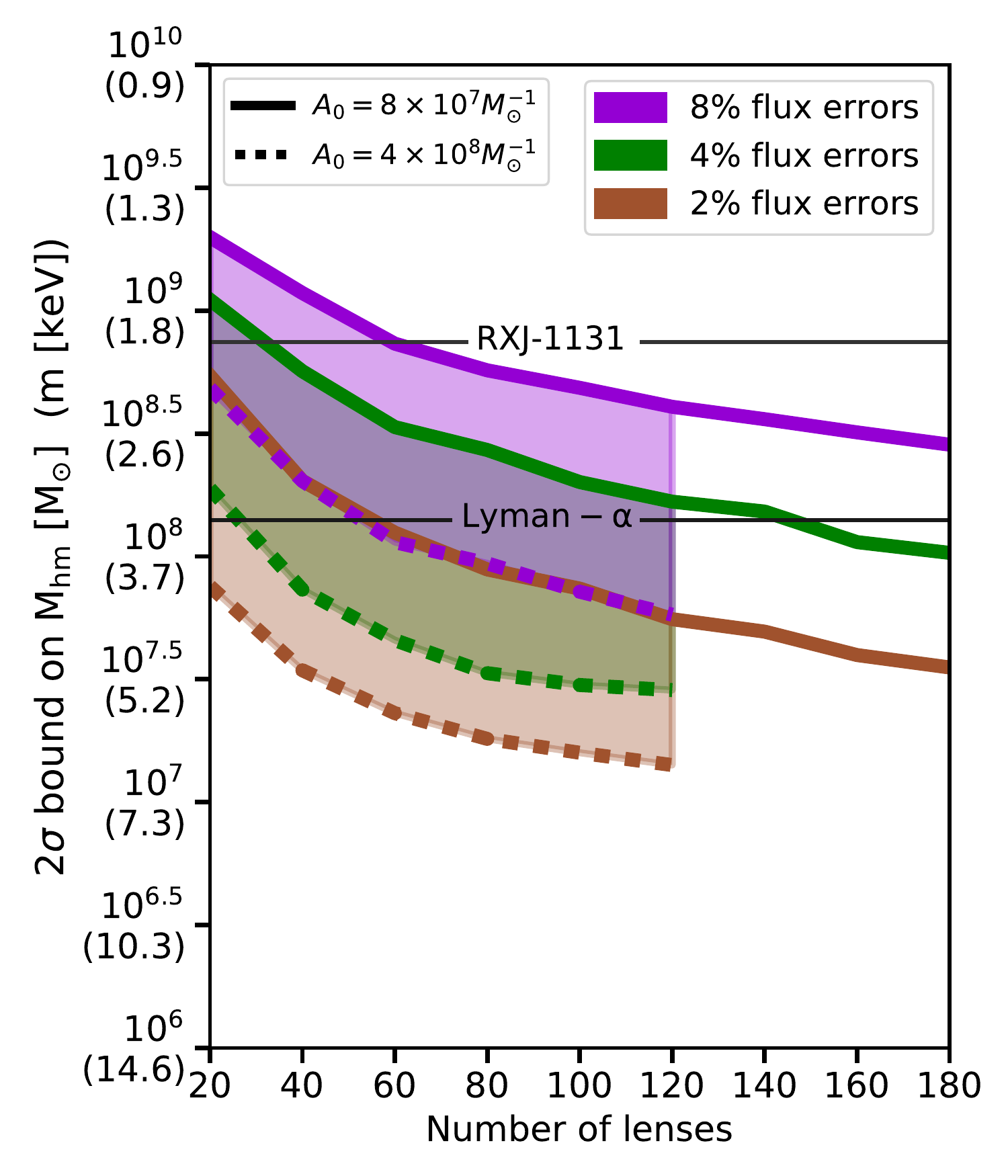}
		\caption{\label{fig:nlens_vs_2sigma} Curves show the dependence of the $2\sigma$ bounds on $m_{\rm{hm}}$, and the mass $m$ of a corresponding thermal relic dark matter particle, as a function of the number of lenses. For each plotted point, we randomly sample different combinations of $N$ lenses, each with a random draw of flux errors of 2, 4, and 8 percent. We iterate this procedure 200 times, and compute the mean of the two sigma bounds over the 200 iterations . 
			With 180 lenses, the $2\sigma$ bound on $m_{\rm{hm}}$ is $10^{6.4} \msun$, $10^{7.5} \msun$, $10^{8} \msun$, $10^{8.4} \msun$ for fluxes with errors controlled at $0\%$, $2\%$, $4\%$, and $8\%$, respectively. For the higher normalization case, (dotted curves) the signal to noise ratio in the data is higher, allowing tighter constraints for a fixed sample size. For reference, horizontal lines show bounds on the mass of a thermal from the Lyman-$\alpha$ forest \citep{Viel13} and an analysis of the strong lens RXJ-1131 \citep{Birrer++17}.}
	\end{figure}
	
	In this Subsection we demonstrate the effect of flux errors on our inference.
	
	We start by using the the same data drawn from the low subhalo mass normalization scenario as is plotted in Figure \ref{fig:cdm_inf_noerror}. We add flux ratio errors of 4 and 8 percent to the data, and add errors of the same form to the forward model. Random flux errors applied to the data and forward model weaken the constraint on $A_0$  by a factor of $\sim2$ and on $ m_{\rm{hm}}$ over an order of magnitude, as shown in Figure \ref{fig:cdm_inf_error}. This is in part due to the loss of signal from subhalos below masses $10^8 \msun$. The $2\sigma$ posterior probability contours slope upwards, mirroring the degeneracy seen in Figure \ref{fig:wdm_inf_noerror}. We also explore the effect of flux errors in the case of a higher normalization (Figure \ref{fig:cdm_inf_error_highnorm}). A higher normalization results in more frequent high-significance flux perturbations from substructure, which translates to stronger constraints on $m_{\rm{hm}}$ for a fixed number of lenses, and for fixed flux uncertainties.
	
	In the case of low normalization, the inference with $4\%$ errors rules out models with no substructure and very warm mass functions, but with this specific realization of flux errors applied to the data does not quite surpass the bounds set by \citet{Viel13} with the Lyman-$\alpha$ forest. If more subhalos survive in the lens plane than the most extreme models for tidal stripping suggest, the constraints on the free streaming length improve to $\approx 5.5$ keV, as shown in Figure \ref{fig:cdm_inf_error_highnorm}. Of course neither of these scenarios incorporate the boost to the overall normalization which will comes from line of sight structure, so the low normalization case may be seen as a lower limit on the signal we expect. In a future paper we will consider the effects of this additional signal.
	
	We note that our method can robustly distinguish between the two normalization scenarios even in the case of 8\% flux errors. This is important as it indicates that we are able to measure the difference between WDM and a high normalization and CDM and a low normalization. This can potentially provide useful input to simulations of tidal disruption in massive halos, although the signal from the lens plane will be somewhat diminished by the boost from line-of-sight structure.
	
	Figures \ref{fig:cdm_inf_error} and \ref{fig:cdm_inf_error_highnorm} each show one realization of flux perturbations applied to the data. In Figure \ref{fig:nlens_vs_2sigma}, we average over many realizations of flux perturbations to properly account for the constraints possible from $N$ lenses. Stepping through samples of $N$ lenses in increments of 20, we compute 200 bootstraps for each sample, and repeat this procedure for errors of $2\%$, $4\%$, and $8\%$. The resulting curves in reflect a compromise between more precise flux measurements and increasing the lens sample size. Both options improve the constraints afforded by image flux ratios on the subhalo mass function, but increasing the flux precision by better handling systematic errors in the lens modeling and by more accurate measurements results yields better marginal gains. The curves do not appear to level off, suggesting that these bounds may improve by including more lenses in the inference, but without simulating more systems it is difficult to make definitive statements for the regime $N>180$. The normalization is also observed to play a key role, as it effectively boosts the signal to noise ratio in the data and enables measurements more robust to deviations in the fluxes at the percent level.
	
	Conceptually, one can interpret the dependence of the $2 \sigma$ bounds on different flux errors $\delta {F}$ as tracers of the probability distribution $p\left(\delta F | \msub\right)$. Supposing that $m_{\rm{s}}*$ defines the subhalo mass scale that dominates this probability density, it follows that adding random noise at the $\delta F$ level will erode the sensitivity to mass scales $\approx m_{\rm{s}}*$, so the fast deterioration of constraining power on $m_{\rm{hm}}$ tracks the loss of sensitivity to mass scales $\approx m_{\rm{s}}*$. In this context, a higher normalization increases the probability at each scale $m_{\rm{s}}*$ of observing a flux ratio anomaly $\delta F$, counteracting the loss of signal induced by adding flux errors at the $\delta F$ level.
	
	\section{Discussion and Conclusions}
	
	We have introduced a new method to infer the nature of dark matter from observations of flux ratios in quadruply lensed quasars. The method uses an Approximate Bayesian Computing algorithm to statistically infer the input parameters describing the subhalo mass function without directly computing a likelihood function.  We have illustrated the method by performing simulations of strong lenses systems with substructure populations of NFW subhalos rendered in the lens plane, in the case of cold and warm dark matter, and for two different normalizations of the subhalo mass function. 
	
	While a real sample of lenses will be diverse in both redshift distribution and host halo mass, the lens and source redshift will primarily impact the contribution from the line of sight, while the connection between the normalization and host halo mass, and the effect on lensing observables, is subject to considerable theoretical uncertainty \citep{DespVeg16,GK++17}. We handle this uncertainty by considering two limiting cases of the normalization, corresponding to scenarios with and without complete subhalo disruption within the scale radius of the parent halo. As the normalization of each lens effectively weights the information content available, the limiting cases we analyze bound the constraints from a sample of lenses with diverse halo masses.
	
	Our main results can be summarized as follows:
	
	\begin{itemize}
		\item In an idealized scenario, where the macromodel is known to high precision and other sources of flux ratio errors are mitigated, the only source of flux ratio perturbation comes from dark substructure and flux ratios probe the mass function at scales below $10^7 \msun$. With flux uncertainty at the level $2\%$, $4\%$ and $8\%$, the bounds on the half-mode mass [$M_{\odot}$] (thermal relic mass [$\rm{keV}$]) are $m_{\rm{hm}}=10^{7.5} \ (5.0)$, $m_{\rm{hm}}=10^{8} \ (3.6)$ and $m_{\rm{hm}}=10^{8.4} \ (2.7)$ with 180 systems. For the higher normalization case, the improvement in the signal to noise ratio in the data yields constraints of $m_{\rm{hm}}=10^{7.2} \ (6.6)$, $m_{\rm{hm}}=10^{7.5} \ (5.3)$ and $m_{\rm{hm}}=10^{7.8} \ (4.3)$ with just 120 lenses. In a WDM scenario, we find with no uncertainty in flux ratios that we can measure the position of the free-streaming cutoff in the subhalo mass function with just 30 lenses, constraining it to between $10^{7}M_{\odot}$ and $10^{8.7} M_{\odot}$ at $2 \sigma$. With less control over systematics and degraded measurement precision, more than 30 lenses would be required to achieve these constraints, but our simulations suggest that this method can, in principle, measure the warmth of dark matter should CDM be the incorrect model. Provided one controls for systematic errors in flux ratios associated with incorrect macromodels, these constraints will likely improve after adding the contribution from line of sight structure, which contributes substantial additional signal.
		\item The $2 \sigma$ bound on $m_{\rm{hm}}$ improves rapidly with increasing flux uncertainties, and falls slowly after $N \approx 100$ lenses. This reflects the sensitivity of flux ratios to low mass subhalos, which impart deviations at the level of a few percent. In terms of overall strategy for the study of strong lens systems, this establishes the necessity of measuring fluxes precisely and controlling for systematic errors arising from the parameterization of the macromodel. The simple SIE+shear parameterization implemented in this work may not be sufficient for systems in which additional mass components are present in the main deflector, such as stellar disks. In practice, identifying morphological complexity in the main deflector can be achieved by deep imaging of the lensing galaxy to identify luminous mass components, and preferentially analyzing slow-rotators with high central velocity dispersions. 
		\item The frequency and magnitude of flux ratio anomalies differentiates between different dark matter models. In this work, we have explored the effect of a varying normalization and half-mode mass. While the half-mode mass scale carries information regarding the nature of dark matter, baryonic effects can impact the normalization, which plays a crucial role as it effectively determines the signal to noise ratio in the data. This translates into better constraints on the shape of the mass function, making it easier to distinguish WDM from CDM. There is some degeneracy between the normalization of the mass function and the half-mode mass, but flux ratios are sensitive enough to break this degeneracy and probe mass scales below $10^8 \msun$, where CDM and WDM subhalo abundance differs significantly. 
	\end{itemize}
	
	Recent analysis has shown the contribution from line-of-sight subhalos is substantial \citep{Despali++17}. Since inclusion of the line-of-sight structure will likely improve our projected constraints, we interpret our results as understated limits on the power of substructure lensing. We leave the extension of our method to include line-of-sight structure to future work. 
	
	Comparing the posterior probability distributions in Figures \ref{fig:cdm_inf_error} and \ref{fig:cdm_inf_error_highnorm}, although there is some degeneracy between the normalization and the half-mode mass these parameters can be constrained simultaneously with $\approx 100$ lenses. \footnote{In fact, the differences between the low normalization and high normalization scenarios become apparent with fewer than 100 lenses, but the extent to which one can differentiate the two depends on the degree to which one controls for systematic errors in flux ratios, and the difference between the two normalizations.} The normalization has important implications for dark matter and baryonic physics through its connection to total halo mass and tidal stripping of subhalos, respectively, and thus potentially will provide an important constraint for theoretical models. 
	
	Our method hinges on accurate measurement of image flux ratios and controlling for systematic errors in their modeling. To quantify the impact of small uncertainties in these observables, we simulate observations with different errors applied to the image fluxes to erase information at the $2\%$, $4\%$, and $8\%$ level, and find the projected constraints are extremely sensitive to loss of information at this level. Case in point: the difference between perfect models and perfect measurements, and an observational scenario with $2\%$ uncertainties in image fluxes is an order of magnitude in $m_{\rm{hm}}$. To achieve the required level of measurement precision, we will need flux ratios computed from the narrow-line emission of the background quasar \citep{Nie++14,Nierenberg++17}, which yield measurements accurate to $4-6\%$ in flux, and are resilient to microlensing. The presence of systematic errors in the modeling can also be mitigated by restricting analysis to deflectors with high velocity dispersions and no complicated morphological features like stellar disks, assuring that dark substructure dominates as the source of flux ratio anomaly computed with respect to simple SIE macromodels. Alternatively, deep imaging of the deflector and its stellar mass distribution may enable the construction of lens models that map the luminous structure of the deflector in detail, as was done in \citet{HsuehEtal16,Hsueh++17}. For these more complicated systems, additional observable information in the form of deep imaging is required to constrain the mass distribution of the deflector.
	
	Finally, we note that our method accommodates any arbitrary dark matter model, provided it specifies the form of the subhalo mass function and the density profiles of individual substructures. Possible extensions of our method can explore subhalo populations from mixed or self-interacting dark matter \citep{Rocha++13,Todoroki17}, and models in which a fraction of the dark matter is composed of primordial black holes \citep[e.g.][]{CotnerKus17}. 
	
	\section*{Acknowledgments}
	We thank Adriano Agnello, Andrew Benson, Michael Boylan-Kolchin, Francis-Yan Cyr-Racine, Chris Fassnacht, Stacy Kim, Alex Kusenko, Phil Marshall, Leonidas Moustakas, Annika Peter, Simona Vegetti, and Dandan Xu for helpful suggestions and interesting discussions throughout the course of this project. We also thank the anonymous referee for feedback that improved the quality of the paper.  
	
	DG, TT, and SB acknowledge support by the US National Science Foundation through grant AST-1714953. CRK acknowledges support by the NSF through grant AST-1716585. This work used computational and storage services associated with the Hoffman2 Shared Cluster provided by the UCLA Institute for Digital Research and Education's Research Technology Group. 
	
	\bibliographystyle{mnras}
	\bibliography{references}
	
	\appendix
	
	\section{\bf More on generating substructure realizations}
	\label{app:A}
	\subsection{Spatial distribution and truncation}
	As discussed in Section \ref{sec:sim_setup}, we tidally truncate the subhalos according to their 3-d position in the halo $r_{\rm{3d}}$ according to
	\begin{equation} \label{eqn:truncrad2}
	r_t = \left(\frac{m_{s} r_{\rm{3d}}^2}{2 \Sigma_{\rm{crit}} R_{\rm{Ein}}}\right)^{\frac{1}{3}}
	\end{equation}
	\cite[see][]{CyrRacine++16}, where $m_s$ refers to $M_{200}$ at redshift $z_d = 0.5$. Although strong lensing quantities typically live in projection in the lens plane (omitting the line of sight), the third spatial dimension enters through the truncation, via the 3-d position $r_{3d} = \sqrt{z^2+r_{2d}^2}$.  
	
	To include this effect in our simulations, we begin by noting that in projection, in the inner portions of a galactic halo where strong lensing takes place, subhalos appear distributed uniformly in two dimensions \citep{Xuetal2015}. We therefore assign each subhalo a projected position $r_{2d}$ with a spatially uniform probability density out to $R_{\rm{max}} = 18.6 \rm{kpc}$, or 3 arcseconds at the lens redshift.
	
	To obtain the 3-dimensional z coordinate for a subhalo, we start with a two dimensional distribution that is uniform (to a very good approximation) with in $18.6 \ \rm{kpc}$
	\begin{equation}
	p\left(r_{2d}|r_c\right) \propto \left(1+\frac{r_{2d}^2}{r_c^2}\right)^{-1}
	\end{equation}
	with $r_c = 75 \rm{kpc}$. We then de-project this 2-d density into the third dimension to obtain a density for the z coordinate, out to a maximum 3-d radius $R = 250 \rm{kpc}$. The corresponding distribution for the z coordinate, given a 2-d position, becomes
	
	\begin{equation}
	p\left(z | r_{2d},r_c\right) \propto \left(1+\frac{z^2+r_{2d}^2}{r_{c}^2}\right)^{-1.5}.
	\end{equation}
	We note that this has the same asymptotic form $\propto r^{-3}$ as an NFW profile for large z. 
	The z coordinate affects the lensing only indirectly through the truncation radius. For an NFW profile, this does not significantly impact the image magnification, as this observable is principally determined by the central density which is unchanged. We verify that our truncation scheme consistently yields $r_t > r_s$. Typical values for $\frac{r_t}{r_s}$ range between 5-30, depending on the concentration of the subhalo. 
	
	\subsection{Mass Function}
	
	Focusing first on the form of the mass function $\frac{dN}{dm}$, numerical simulations of cold dark matter halos \citep{Spr++08,Gao++12,Fiacconi++16} suggest a scale free mass function $\frac{dN}{dm_s} \propto m_s^{-1.9}$ for $10^6 \leq \frac{m_s}{M_{\odot}} \leq 10^{10}$. We acknowledge that tidal disruption may alter this prediction significantly in the inner portions of galactic halos, but we do not address this concern here, as our principle aim is to demonstrate the method rather than focus on the most realist mass function. 
	
	When normalizing the subhalo mass function, we wish to compare CDM realizations with WDM realizations with the same amplitude at mass scales far above the half-mode mass $m_{\rm{hm}}$. To accomplish this, we start with a scale free CDM mass function 
	\begin{equation}
	\label{eqn:CDM_mfunc}
	\frac{dN_{\rm{cdm}}}{dm_{{s}}} = A_0 \left(\frac{m_s}{\msun}\right)^{-\alpha}
	\end{equation}
	taking $\alpha = 1.9$. In the regions of dark matter halos, the spatial distribution of substructure is approximately uniform in projection \citep{Xuetal2015}. Uniformly distributing subhalos in a plane with radius $R_{\rm{max}}$, we relate the substructure convergence at the Einstein radius, $\kappa_{\rm{sub}}$, to the mass in substructure between $m_L$ and $m_{\rm{s}} \leq m_H$
	\begin{equation}
	\label{eqn:norm}
	\Sigma_{\rm{crit}} \kappa_{\rm{sub}} \pi R_{\rm{max}}^2 = \int_{m_L}^{m_H} m_s \frac{dN_{\rm{cdm}}}{dm_s} dm_s.
	\end{equation}
	This yields the normalization $A_0$ in terms of $\kappa_{\rm{sub}}$, and the mean number of subhalos
	\begin{eqnarray} \label{eqn:norm_params}
	\nonumber A_0 &=& \frac{\left(2-\alpha \right)\pi R_{\rm{max}}^2 \Sigma_{\rm{crit}} \ \kappa_{\rm{sub}}}{M_{\sun}^{\alpha}\left(m_H^{2-\alpha} - m_L^{2-\alpha}\right)} \\ 
	\langle N_{s} \rangle &=& \frac{A_0}{1-\alpha} \left(m_H^{1-\alpha} - m_L^{1-\alpha}\right) \msun^{\alpha}.
	\end{eqnarray}
	We then draw $N_s$ subhalos from a Poisson distribution with average value $\langle N_{s} \rangle$.
	
	\subsection{Normalization}
	
	The normalization of the subhalo mass function depends both on the accretion history and evolution of subhalos in the lens halo, and the effects of baryonic physics in the central regions of the halo. The former effect, the accretion history of dark matter halos as a function of halo mass, has been well studied, and here we adopt the result of \citep{Han++16} and assume a total surviving halo mass fraction of $f_{sub, halo} \sim 6\%$. There is scatter in the predictions depending on halo accretion history and redshift which can raise this value by as much as a factor of 2 \citep[e.g.][]{Fiacconi++16,JiangBosch++17}, however we conservatively adopt the lower normalization for this estimate, and lower the overall subhalo mass normalization for both the extreme and no tidal disruption cases by $30\%$, bringing the assumed total halo mass fraction in substructure to $\sim 4\%$.
	
	The effect of baryonic physics is more uncertain. Recent state of the art hydrodynamic simulations indicate that in a Milky Way mass host halo, the central disk may destroy all subhalos within the central 20 kpc of the host, in addition to reducing the total number of subhalos by $\sim30\%$ compared to a dark matter only run \citep{GK++17}. In these simulations, the destruction appears independent of subhalo mass.
	
	In order to bracket the range of possibilities and to demonstrate how tidal disruption in the central region would affect our inference, we consider two scenarios. In the first, we assume that dark matter subhalos follow the NFW density profile of the host \citep{Han++16}, which is seen in dark matter only simulations. In the second case we mimic the effects of central tidal disruption by assuming that the destruction radius scales with the host scale radius. Dedicated simulations of lens mass halos will be necessary to calibrate this effect in more detail.
	
	For a lower bound on the normalization, we assume that all tidal disruption destroys all subhalos within the scale radius of the host (for a typical lens halo of mass $\sim 10^{13}M_{\odot}$ this radius is $\sim 150$ kpc). For an upper bound, we assume that the subhalo number density follows the density profile of the host. In each case, we compute the projected mass density in substructure along the longitudinal virial radius of the host inside a projected cylinder of 18 kpc ($\sim$ 3 arcseconds at $z = 0.5$), and obtain $\left(\kappa_{\rm{sub}}\right) = 0.002 \ (0.01)$ in the case of extreme (minimal) tidal stripping. These values correspond to values of $A_0 =   8 \ (40) \times 10^7 \msun^{-1}$.
	
	\subsection{Extension to WDM}
	The half mode mass $m_{\rm{hm}}$ corresponds to a characteristic length scale at which the linear matter power spectrum with pure WDM is damped with respect to that of CDM by one-half. For details, see e.g. \citet{Viel++05,Schneider++12}. We wish to compare a range of WDM subhalo populations with varying $m_{\rm{hm}}$ according to the mass function
	\begin{equation}
	\label{eqn:app_wdm_massfunc}
	\frac{dN_{\rm{wdm}}}{dm_s} = \frac{dN_{\rm{cdm}}}{dm_s} \left(1+\frac{m_{\rm{hm}}}{m_s}\right)^{-1.3}
	\end{equation}
	while preserving the amplitude of the mass function for masses high above $m_{\rm{hm}}$ to isolate the effect of $m_{\rm{hm}}$ from the normalization $A_0$. To do this, we first generate subhalos according to Equation \ref{eqn:norm}. This results in scale free mass function, which we deplete by removing subhalos probabilistically with probability
	
	\begin{equation}
	P \propto \left(1+\frac{m_{\rm{hm}}}{m_{\rm{s}}}\right)^{-1.3}.
	\end{equation} 
	This yields substructure populations obeying the curves plotted in Figure \ref{fig:mass_function}. 
	
	\section{\bf Sensitivity of cusp, fold, and cross configurations}
	\label{app:B}
	The response of an image magnification to small scale structure is heightened if it lies close to a critical curve. Similarly, images close to one another may be affected by the same substructures, introducing a correlation between flux anomalies in different images. 
	
\begin{figure}
		\includegraphics[clip,trim=.15cm .5cm 0cm
		0cm,width=.46\textwidth,keepaspectratio]{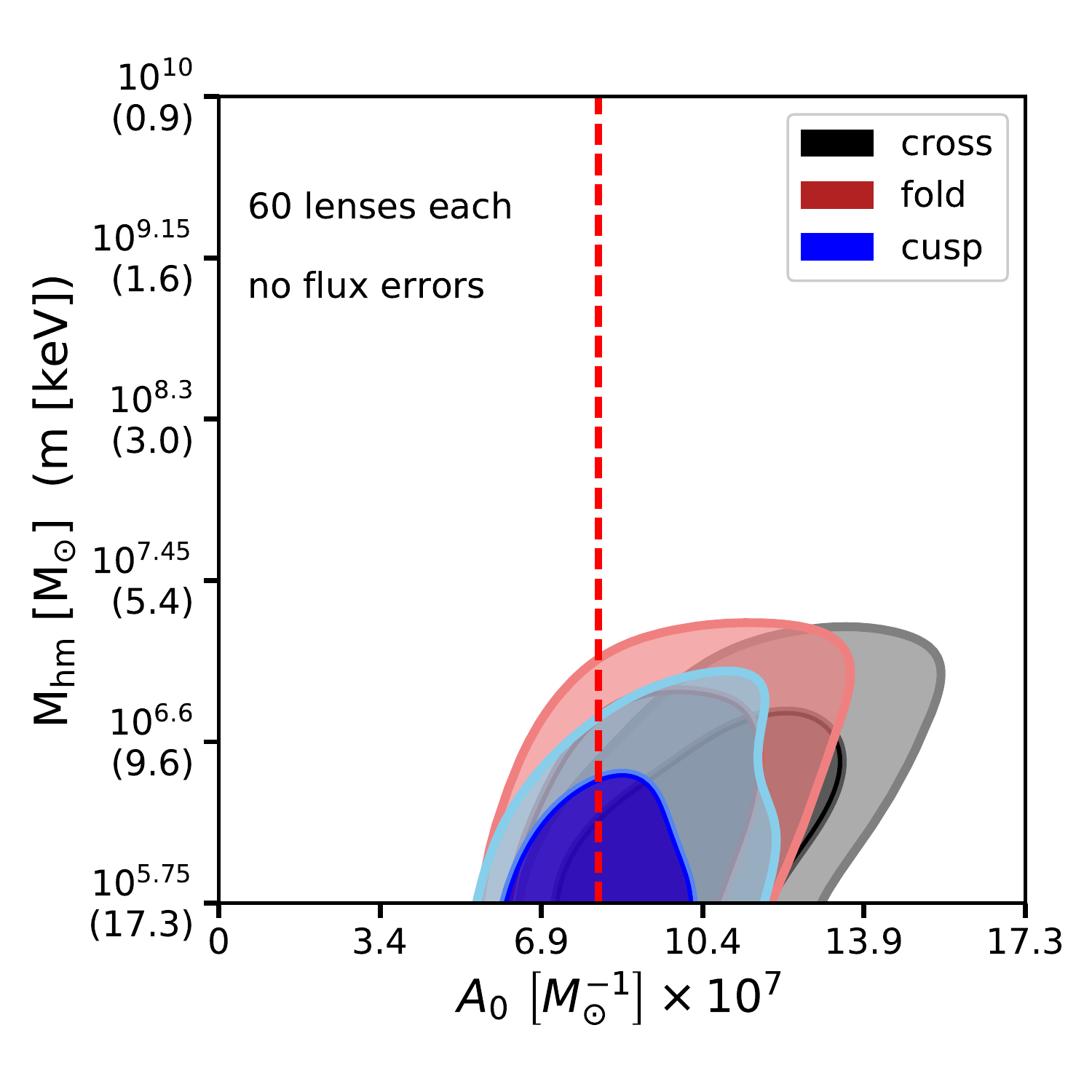}
		\caption{\label{fig:img_configs} Response of cusp, fold and cross image configurations to small scale structure. For a fixed number of lenses, cusps yield the strongest constraints, followed by folds and crosses.}
	\end{figure}

	Cusp (fold) configurations are characterized by three (two) images straddling the critical curve, and by the three (two) images close in proximity to each other. One therefore expects a ranking in sensitivity to substructure of cusp, fold, cross, in descending order. In Figure \ref{fig:img_configs}, we show that this is indeed the case. Interestingly, the degeneracy between a warm mass function with a high normalization, and a cold mass function with low normalization is reduced in cusp and fold configurations compared to crosses. 
	
	To identify image configurations, we adopt the following classification scheme based on the Einstein radius $R_{\rm{Ein}}$ and the image separations. If the smallest image separation is greater than 0.7$R_{\rm{Ein}}$, the lens is immediately classified as a cross. If the second largest separation is $< 1.2R_{\rm{Ein}}$, we classify it as a cusp, and otherwise it is a fold. 
	
	\section{\bf Use of other summary statistics}
	\label{app:C}
	The summary statistic in Equation \ref{eqn:summary_stat} is closely the likelihood, or a $\chi^2$ value. It penalizes models which do not reproduce the anomalies observed in the data in the correct images. There is, however, no `correct' choice of summary statistic, only ones that perform better than others.
	
	We experimented with using other statistics, including $R_{\rm{cusp}}$ and $R_{\rm{fold}}$ \citep[see][]{KGP03,Keeton:2005p591}. An advantageous feature of these parameters is that they can be computed directly from the observed image fluxes, and do not rely on a lens model to identify anomalies. After experimenting with other summary statistics, however, we find the summary statistic in Equation \ref{eqn:summary_stat} yields the strongest constraints, because it does not discard information by adding and subtracting fluxes from different images. 
	
\begin{figure}
		\includegraphics[clip,trim=.15cm .5cm 0cm
		0cm,width=.46\textwidth,keepaspectratio]{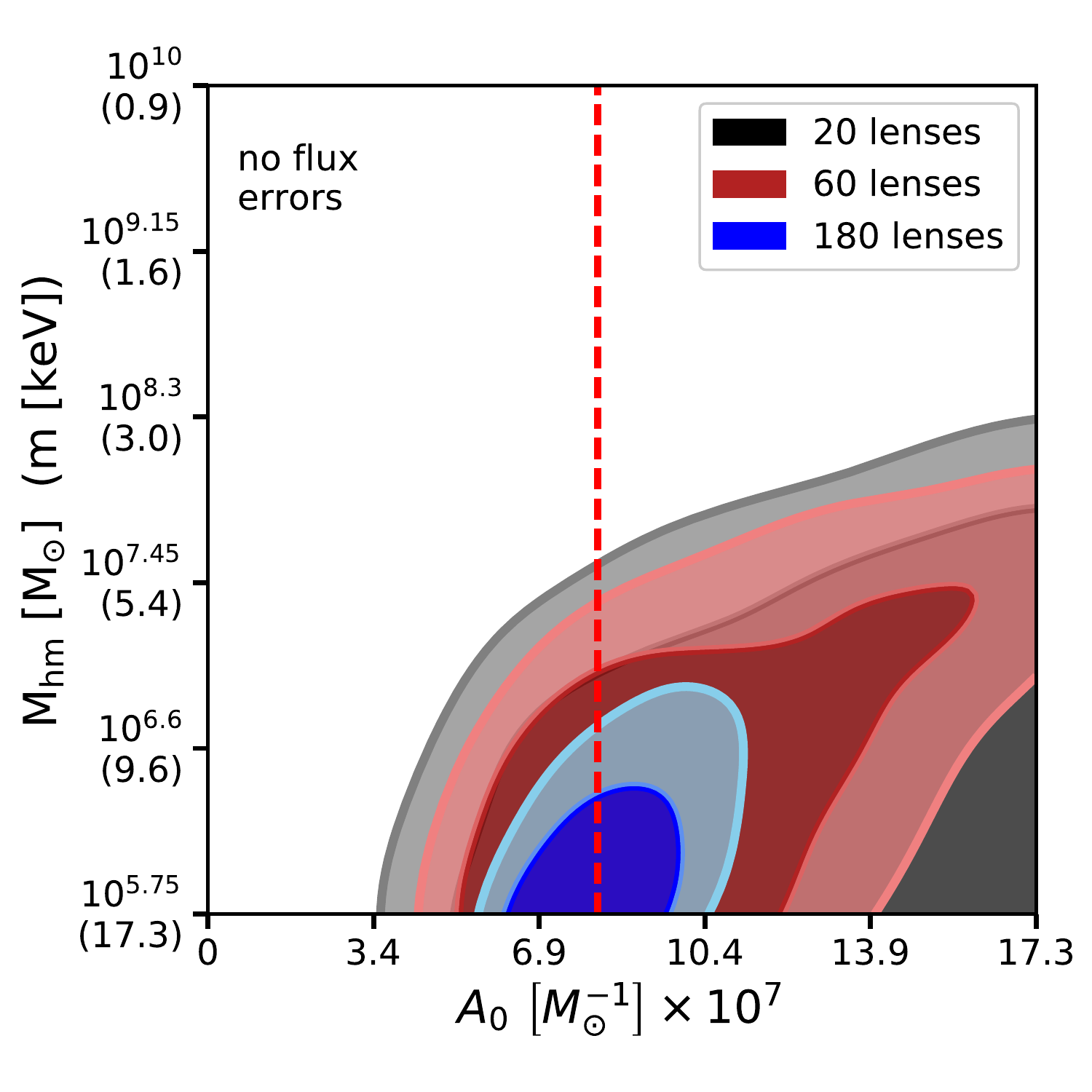}
		\caption{\label{fig:other_stat} Inference made using the statistic in defined in Equation \ref{eqn:stat2}, rather than Equation \ref{eqn:summary_stat}.}
	\end{figure}

	As an example, consider the summary statistic
	\begin{equation}
	\label{eqn:stat2}
	S \left(\fobs,\fmods \right) = \left| \sum_{i=1}^{3} \left(\fobsi - \fmodsi\right) \right| .
	\end{equation} 
	Using this equation instead of Equation \ref{eqn:summary_stat}, we perform an inference on the same simulation as in Figure \ref{fig:cdm_inf_noerror}. The results are shown in Figure \ref{fig:other_stat}. Equation \ref{eqn:summary_stat} performs better. 
	
	\section{\bf Convergence of ABC simulations and posteriors}
	\label{app:D}
	\begin{figure}
		\includegraphics[clip,trim=0cm 0cm 0cm
		0cm,width=.48\textwidth,keepaspectratio]{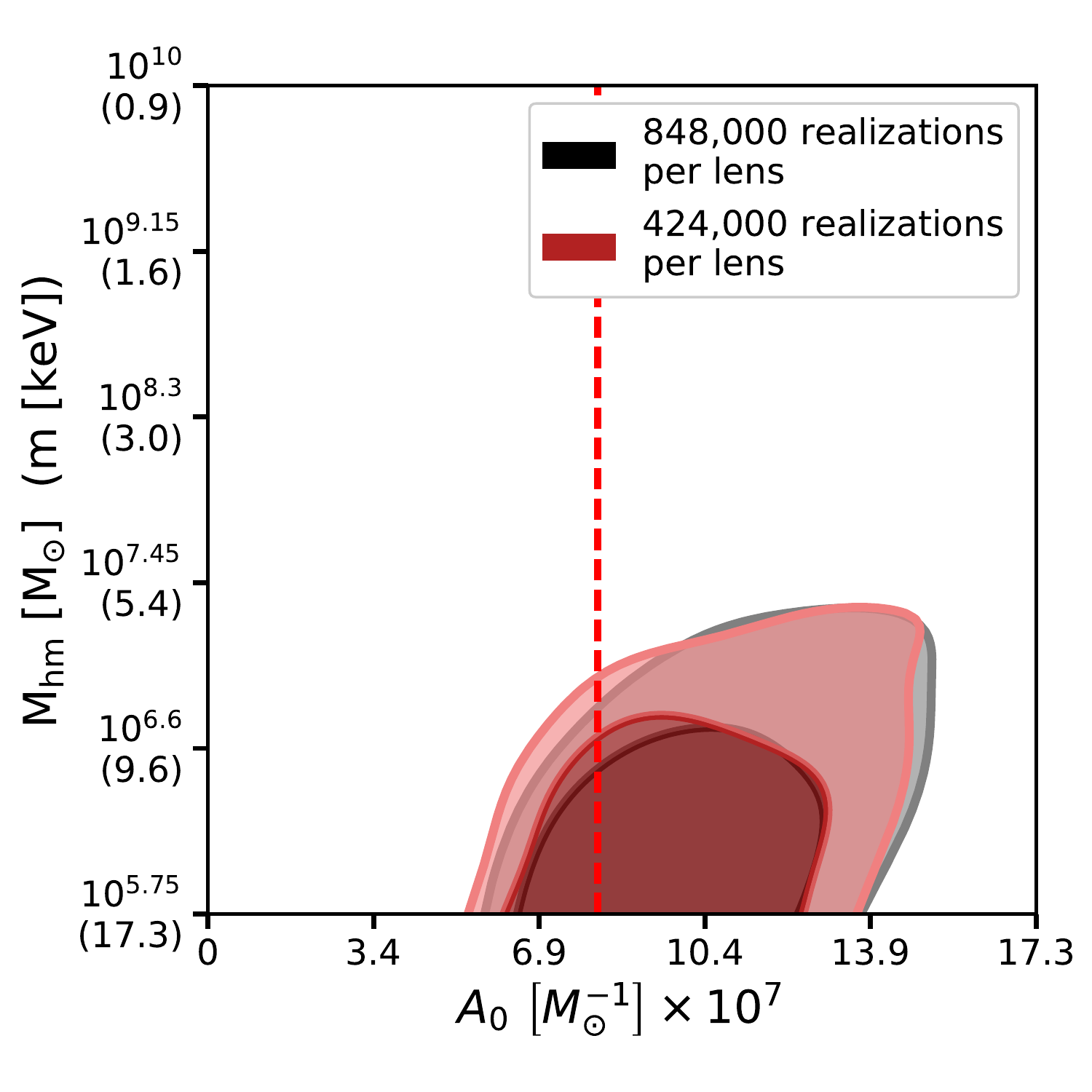}
		\caption{\label{fig:convergence} A convergence test in which we discard half of the realizations before applying the acceptance criterion. Each distribution is composed of the samples from $\qsub$ associated with the lowest 1,800 summary statistics.}
	\end{figure}
	To assess convergence of the ABC algorithm, we compare two inferences made on the same data set in which one has half the number of realizations as the other. To produce the black distribution, we retain the draws from $\qsub$ associated with the lowest 1,800 summary statistics. We then discard half of the realizations, and reject all but the lowest 1,800 summary statistics from the depleted simulation. Under-sampling by a factor of two, we recover the same bounds on $A_0$ and $m_{\rm{hm}}$ to a high degree of precision, indicating the inference made with $\approx 2000$ realizations per proposal $\qsub$ has converged.
	
	\section{\bf Fitting SIE macromodels to datasets built with non-isothermal power laws}
	\label{app:E}
	\begin{figure}
		\includegraphics[clip,trim=0cm 0cm 0cm
		0cm,width=.48\textwidth,keepaspectratio]{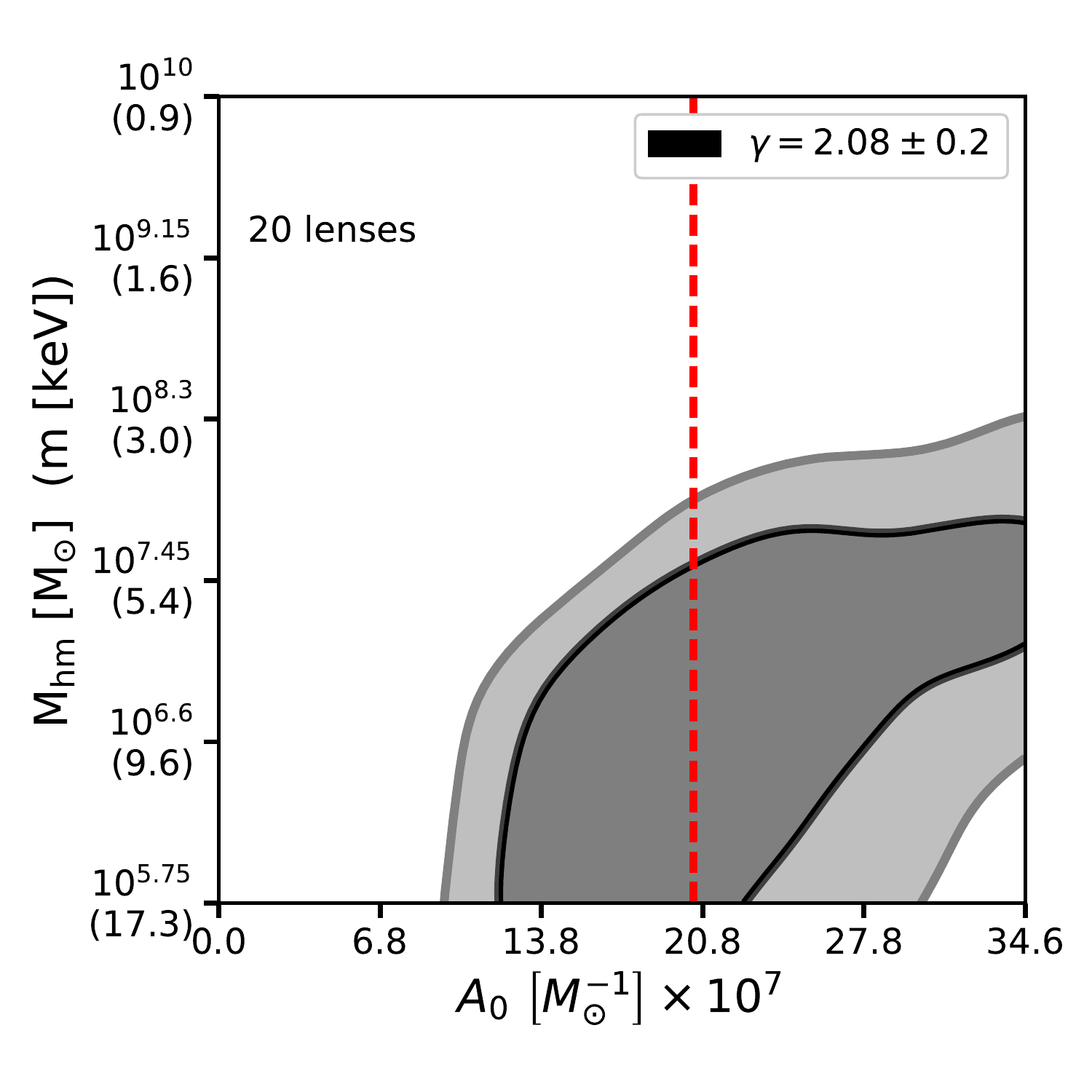}
		\caption{\label{fig:plaw_slope} We test the effect of an incorrectly parameterizing the macromodel by generating mock data with non-isothermal ellipsoids, and modeling the data with an SIE in the forward model. The power law slope $\gamma$ is sampled from a Guassian with mean 2.08 and variance 0.2.}
	\end{figure}
	We test the effect of incorrectly parameterizing the macromodel by created datasets with non-isothermal power law ellipsoids, and fitting them with SIEs in the forward model. When generating mock lenses, we sample power law slopes from a Guassian distribution offset from isothermal at $\gamma = 2.08 \pm 0.2$, consistent with the findings of \citet{Shankar++17}, who find the power law slopes for massive ellipticals are consistently steeper than isothermal. The results of this exercise are shown in Figure \ref{fig:plaw_slope}. The effect of varying the power law slope in the data is seen to not significantly degrade the inference on $A_0$ and $m_{\rm{hm}}$, and therefore does not affect the precision of our forecast statements. 
	
	By completely neglecting the presence of this systematic, we exaggerate its potential bias in the inference. In practice, this systematic should be properly dealt with by sampling different power laws slopes in the forward model. A prior $\gamma$ may be constructed on a lens by lens basis based on measurements of the central velocity dispersion, if available. 
	
\end{document}